\documentclass[twocolumn,amsmath,amssymb,aps,prx,showpacs,longbibliography]{revtex4-1}

\usepackage{graphicx}
\usepackage{epsfig}
\usepackage{dcolumn}
\usepackage{listings}
\usepackage{lipsum}
\usepackage{bm}
\usepackage{hyperref}
\hypersetup{
pdfnewwindow=true, colorlinks=true,
linkcolor=blue, anchorcolor=blue,
citecolor=blue, filecolor=blue,
menucolor=blue, urlcolor=blue}

\usepackage[usenames,dvipsnames]{color}

\newcommand{\unit}[1]{\ensuremath{\, \mathrm{#1}}}
\newcommand{\sub}[1]{\raisebox{-.4ex}{\scriptsize #1}}

\DeclareMathOperator{\tr}{Tr}
\let\Im\undefined
\DeclareMathOperator{\Im}{Im}

\renewcommand{\vec}{\mathbf}
\renewcommand{\mod}{\ensuremath{\, \text{ mod }}}
\begin{document}

\title{Z2Pack: Numerical implementation of hybrid Wannier centers for identifying topological materials}

\newcommand{\lint}{\mathlarger{\int}}
\newcommand{\R}{\mathbb{R}}
\newcommand{\C}{\mathbb{C}}
\newcommand{\N}{\mathbb{N}}
\newcommand{\Z}{\mathbb{Z}}
\newcommand{\Q}{\mathbb{Q}}
\newcommand{\linspan}{\operatorname{span}}
\newcommand{\nextpar}{\vspace{5pt}}

% A command for inner product and bras and kets
\newcommand{\braket}[2]{\left\langle #1 \middle| #2 \right\rangle}
\newcommand{\ketbra}[2]{| #1 \rangle\langle #2 |}
\newcommand{\bra}[1]{\left\langle#1\right|}
\newcommand{\ket}[1]{\left|#1\right\rangle}
\newcommand{\bigket}[1]{\bigl|#1\bigr\rangle}
\newcommand{\textket}[1]{|#1\rangle}

\author{Dominik Gresch$^{1}$}
%\email{greschd@phys.ethz.ch}
\author{Gabriel Aut\`es$^{2, 3}$}
\author{Oleg V. Yazyev$^{2, 3}$}
\author{Matthias Troyer$^{1}$}
\author{David Vanderbilt$^{4}$} 
\author{B. Andrei Bernevig$^{5}$}
\author{Alexey A. Soluyanov$^{1,6}$}
%\email{soluyanov@itp.phys.ethz.ch}
\affiliation{${^1}$Theoretical Physics and Station Q Zurich, ETH Zurich, 8093 Zurich, Switzerland}
\affiliation{${^2}$Institute of Physics, Ecole Polytechnique F\'ed\'erale de Lausanne (EPFL), CH-1015 Lausanne, Switzerland}
\affiliation{${^3}$National Center for Computational Design and Discovery of Novel Materials MARVEL, Ecole Polytechnique F\'ed\'erale de Lausanne (EPFL), CH-1015 Lausanne, Switzerland}
\affiliation{${^4}$Department of Physics and Astronomy, Rutgers University, Piscataway, New Jersey 08854, USA}
\affiliation{${^5}$Department of Physics, Princeton University, Princeton, New Jersey 08544, USA}
\affiliation{$^{6}$Department of Physics, St. Petersburg State University, St. Petersburg, 199034 Russia}

\begin{abstract}
The intense theoretical and experimental interest in topological insulators and semimetals has established band structure topology as a fundamental material property. Consequently, identifying band topologies has become an important, but often challenging problem, with no exhaustive solution at the present time. In this work we compile a series of techniques, some previously known, that allow for a solution to this problem for a large set of the possible band topologies. The method is based on tracking hybrid Wannier charge centers computed for relevant Bloch states, and it works at all levels of materials modeling: continuous $\vec{k} \cdot \vec{p}$ models, tight-binding models and {\it ab initio} calculations. We apply the method to compute and identify Chern, $\mathbb{Z}_2$ and crystalline topological insulators, as well as topological semimetal phases, using real material examples. Moreover, we provide a numerical implementation of this technique (the Z2Pack software package) that is ideally suited for high-throughput screening of materials databases for compounds with non-trivial topologies. We expect that our work will allow researchers to: (a) identify topological materials optimal for experimental probes, (b) classify existing compounds and (c) reveal materials that host novel, not yet described, topological states. 
\end{abstract}
\maketitle

% %%%%%%%%%%%%%%%%%%%%%%%%%%%%%%%%%%%
% \marginparwidth 2.7in
% \marginparsep 0.5in
% \def\dvm#1{\marginpar{\small DV: #1}}
% \def\asm#1{\marginpar{\small AS: #1}}
% \def\scr{\scriptsize}
% %%%%%%%%%%%%%%%%%%%%%%%%%%%%%%%%%%%

\section{Introduction}
\label{sec:intro}
%
% V1 - original version
Topology studies the properties of geometric objects that are preserved under smooth deformations, and divides these objects accordingly into distinct topological classes.
In the past decade the principles of topology were applied to crystalline solids, where electronic bands can have a topological characterization~\cite{Kane-PRL05-b,Hasan-RMP10, Qi-RMP11, Volovik-book}. For example, in band insulators the occupied bands are separated from the unoccupied ones by an energy gap, and form a well-defined manifold in Hilbert space. Certain geometric properties can be defined for this manifold, giving rise  to a topological classification of band insulators~\cite{Schnyder-PRB08, Kitaev-AIP09, Schnyder-AIP09}, and to the notion of topological insulators\cite{Hasan-RMP10, Qi-RMP11}. The physical equivalent of the mathematical notion of smoothly connectable manifolds in this case is the possibility to adiabatically transform one gapped manifold into another. This means that if two gapped Hamiltonians belong to the same topological class, they can be adiabatically connected without a {\it direct} closure of the band gap.

The topological classification of insulators can be enriched by an additional symmetry constraint on the classified Hamiltonians. In this case, two systems are considered to be topologically equivalent if their Hamiltonians can be adiabatically connected by a path along which the band gap remains open {\it and} the symmetry is preserved. Time-reversal (TR) symmetric~\cite{Kane-PRL05-a, Kane-PRL05-b, Fu-PRL07, Moore-PRB07, Roy-PRB09-a, Roy-PRB09-b}, antiferromagnetic~\cite{Mong-PRB10, Fang-PRB13} and crystalline topological insulators~\cite{Fu-PRL11, Parameswaran-NatPhys13, Alexandradinata-PRB16, Alexandradinata-PRL14, Taherinejad-PRB14} are examples of these symmetry-enriched topological classifications.

Unlike other observables, quantum numbers describing the topology of a state do not necessarily correspond to eigenvalues of some Hermitian operator. Instead, a different type of quantum numbers -- topological invariants -- has to be defined in such a way that a distinct number is assigned to each class. The task of identifying topological states then reduces to defining sensible topological invariants that discern different classes. Finding ways to compute these invariants becomes of major importance in the field. A final predictive theory of all topological invariants for all existing topological classes is missing and, due to the multitude of symmetry space groups and possible orbitals at the Fermi level, seems to be out of reach at the present time. 

The ability to distinguish distinct topological classes is not only of theoretical interest, but also allows for the prediction of physical phenomena in real materials. For example, in two dimensions generic insulators with no symmetries apart from the fundamental charge conservation are classified according to the value of the (first) Chern number $C$~\cite{Nakahara-book}. This is a unique characteristic of the occupied manifold. Insulators with $C\neq 0$, called Chern insulators, realize the integer quantum Hall effect in the absence of  an external magnetic field~\cite{Haldane-PRL88, Volovik-88}, and their Hall conductance is related to the Chern number as $\sigma_{xy} = C e^2/h$~\cite{Thouless-PRL82, Kohmoto-AP85}, where $e$ is the electron charge and $h$ is the Planck constant.

The invariants of symmetry-protected band topologies are usually more complex, giving rise to a variety of physical phenomena such as the existence of topologically protected surface states~\cite{Hasan-RMP10, Qi-RMP11}, quantized magneto-electric response~\cite{Qi-PRB08, Essin-PRL09}, the quantum spin Hall effect~\cite{Kane-PRL05-a, Kane-PRL05-b, Bernevig-Science06, Konig-Science07}, and non-Abelian quasi-particles for topological quantum computing~\cite{Kitaev-Usp01, Fu-PRL08, Lutchyn-PRL10, Oreg-PRL10, Alicea-PRB10, Kitaev-Annals03} in topological superconductors and superfluids~\cite{Volovik-book}. 

Metals also allow for a topological characterization~\cite{Volovik-book}. For indirect band gap semimetals, where the lower-lying bands are gapped from the rest at each momentum in the Brillouin zone (BZ), the topology of the lower lying states is defined in a similar way to that of the occupied states in insulators. At the surfaces of these metals, topologically protected (sometimes discontinuous) surface states (Fermi arcs) can coexist with projected bulk metallic states. For metals with a vanishing direct band gap, topological invariants akin to those of insulators can sometimes still be defined on surfaces/lines in the BZ, on which the bands of interest are gapped. A suitable choice of the chemical potential in a calculation also allows for topological invariants to be defined for metallic Fermi surfaces~\cite{Haldane-PRL04}. Non-trivial topology in semimetals, such as Weyl and Dirac nodes, can significantly affect observed quantities, for instance their electromagnetic response~\cite{Nielsen-PLB83, Burkov-PRL11, Burkov-PRB11, Zyuzin-PRB12, Aji-PRB12, Ali-Nat14, Son-PRB13, Huang-PRX15, Lv-PRX15, Liang-NatMat15, Gosalbez-PRB15}.   

Many real material examples of topological insulators and metals have been discovered so far. They  realize {\it some} of the theoretically predicted topological phases (see for example Refs.~\cite{Bernevig-Science06, Zhang-NatPhys09, Dai-PRB08, Teo-PRB08, Konig-Science07, Chen-Science09, Hsieh-Nature08, Hsieh-Science09, Xia-NatPhys09, Xiao-PRL10, Feng-PRL11, Wada-PRB11, Neupane-NatComm14, Borisenko-PRL14, Wang-PRB13, Liu-NatMat14, Liang-NatMat15, Wang-PRB12, Liu-Sci14, Xu-Science15, Lv-PRX15}). The identification of new candidate materials, better suited for experimental studies and for realizing novel topologies, is a high priority for the field, but again a common, exhaustive, search procedure seems out of reach at the present time. Given the amount of existing materials and current abilities in synthesis and growth of novel compounds and heterostructures, it is desirable to develop a methodology and software that would allow for a routine computation of some of the various known topological invariants in a way accessible to non-specialists.         

In this paper we develop such a general methodology and present a software package -- Z2Pack -- based on it. This software can be used as a postprocessing tool with most existing {\it ab initio} codes, or as a standalone tool for analyzing topological structure of tight-binding or $\vec{k} \cdot \vec{p}$ Hamiltonians. Z2Pack is ideally suited for a high-throughput search of topological materials. It can also be used to design materials or devices with specific topological properties and to identify not only new topological materials, but also novel topological classes thereof.  

Topological invariants of superconductors in the Bogoliubov-de Gennes (BdG) representation can also be studied, since they are also described by tight-binding or $\vec{k} \cdot \vec{p}$ models, supplemented with the particle-hole symmetry. For simple models a specially designed online interface is provided (\url{http://z2pack.ethz.ch/online/}), allowing one to obtain topological invariants without installing the software. The code represents one of the main results of this paper. We hope it will bring the field of topological invariants in realistic materials to every interested researcher.

The method is based on the concept of hybrid Wannier functions (HWF)~\cite{Sgiarovello-PRB01}, which are localized in only one direction, remaining delocalized in the others. It was shown previously~\cite{Coh-PRL09, Soluyanov-PRB11-a, Soluyanov-PRB11-b, Soluyanov-thesis, Yu-PRB11, Alexandradinata-PRB14, Taherinejad-PRB14} that the flow of HWF charge centers reveals the non-trivial topology of Chern~\cite{Haldane-PRL88, Volovik-88} and TR-symmetric~\cite{Kane-PRL05-b, Fu-PRL07} topological insulators and can be used to compute the corresponding topological invariants~\footnote{Interestingly, HWF charge centers were argued to provide a measure of the state topology in optical lattices~\cite{Wang-PRL13, Grusdt-PRA14}. Some of these proposals were later verified experimentally~\cite{Duca_Sci15}.}.

This technique can be generalized to crystalline topological insulators \cite{Alexandradinata-PRB16, Alexandradinata-PRL14, Taherinejad-PRB14} and topological semimetals \cite{Wan-PRB11, Wang-PRB12, Soluyanov-Nat15, Wang-PRL16, Autes-PRL16, Bzdusek-Nat16, Zhu-PRX16, Muechler-ARX16}. For the former, spatial symmetries give rise to non-trivial band topologies. For the latter, the method allows identifying the presence of a topological phase by means of defining and computing various invariants analogous to those of topological insulators. In Weyl (Dirac) semimetals for example, the flow of HWF charge centers on certain surfaces reveals the presence, location and (for Weyl semimetals) chirality of the Weyl (Dirac) points or lines. 
The universality of this method allows us to develop a general strategy for finding topological features in band structures.

The paper is structured as follows: a review of HWFs is given in Sec.~\ref{sec:hwf_wcc}, along with the representation of Chern numbers in terms of HWF charge centers. In Sec.~\ref{sec:strat} a general strategy for identifying topological materials is described. Illustrations for the application of this strategy to Chern, TR-symmetric and crystalline insulators are given in Sec.~\ref{sec:insulators}.  The extension of the HWF technique to topological semimetals is developed and extensively illustrated in Sec.~\ref{sec:semimetals}. Finally, we describe the numerical implementation of the method in Sec.~\ref{sec:impl} and present an outlook along with some concluding remarks in Sec.~\ref{sec:conclusion}. 

\section{Hybrid Wannier charge centers and Chern numbers}
\label{sec:hwf_wcc}

Here we review the definition and basic properties of HWFs and their charge centers. A basic topological invariant -- the (first) Chern number -- is cast in terms of the HWF charge centers. The flow of HWF centers is introduced as the main signature of non-trivial topology.     

\subsection{Hybrid Wannier functions}\label{ssec:hwf}

Electronic states in crystalline solids are most commonly represented with Bloch functions $\psi_{n\vec{k}}(\vec{r})=\langle \vec{r}|\psi_{n\vec{k}}\rangle$, which, according to the Bloch theorem, take the form
\begin{equation}
\psi_{n{\bf k}}({\bf r}) = e^{i {\bf k}\cdot {\bf r}}u_n({\bf r}),
\end{equation}
where $n$ is the band index and
\begin{equation}
u_{n {\bf k}}({\bf r}) = u_{n {\bf k}}({\bf r} + {\bf R})
\end{equation}
is the lattice-periodic part of the wavefunction. Being essentially modulated plane waves, Bloch functions are delocalized in real space.

In many problems, however, the use of a local basis is preferred. This basis is provided by Wannier functions (WFs) $w_n(\vec{r}-\vec{R})=\langle \vec{r}|\vec{R} n\rangle$ that are obtained from the Bloch states by a Fourier transformation
\begin{equation}
|\vec{R} n\rangle = \frac{V}{(2\pi)^d}\int_{\mathrm{BZ}}  e^{-i \vec{k}\cdot \vec{R}} |\psi_{n, \vec{k}}\rangle \mathrm{d}\vec{k}
\end{equation}
where $d$ stands for the space dimensionality, $V$ is the unit cell volume, and the integral is taken over the first BZ. 

Thus defined WFs are not unique. An isolated~\footnote{A set of bands is called isolated, when the energy gap separates it from the bands above and below at each $k$-point.} set of $N$ energy bands corresponding to  the Bloch Hamiltonian eigenstates $|\psi_{n{\bf k}}\rangle$ can equivalently be described by an alternate set of $N$ Bloch wavefunctions that might not be the Hamiltonian eigenstates, but span the same Hilbert space \footnote{This general rotation on the space of occupied eigenstates is an exact symmetry of a ``flat-band'' Hamiltonian, where all occupied energies have been set to a single value.}. That is, a general unitary basis transformation, called gauge transformation, of the form 
\begin{equation}
|\tilde{\psi}_{m \vec{k}}\rangle = \sum_{n=1}^N U_{nm}(\vec{k})|\psi_{n \vec{k}}\rangle
\label{eqn:gauge}
\end{equation}
can be performed prior to constructing WFs for the given set of bands. Depending on the gauge choice, the resultant WFs can have different properties~\cite{Marzari-PRB97}, in particular their shape and localization in real space can differ significantly. The construction of exponentially localized WFs $w_n(\vec{r}-\vec{R})$ requires the gauge to be smooth, meaning that the Bloch states used to construct WFs are smooth and periodic in reciprocal space. We will see below that such a gauge choice is not always possible \cite{Thouless-JPC84, Brouder-PRL07, Panati-AHP07, Panati-CMP13, Fiorenza-AHP16}.

For the purposes of the present paper, the most convenient basis is that of HWFs~\cite{Sgiarovello-PRB01, Soluyanov-PRB11-a}, which are Wannier-like in one direction but Bloch-like in the others. The formal definition is \footnote{w.l.o.g. we choose the x-direction to be Wannier-like and the others to be Bloch-like}
\begin{equation}
|n;\ell_x, k_y, k_z\rangle = \frac{a_x}{2\pi}\int\limits_{-\pi/a_x}^{\pi / a_x} e^{i k_x \ell_x a_x} |\psi_{n\vec{k}}\rangle \mathrm{d}k_x,
\end{equation}
where $\ell_x \in \mathbb{Z}$  and $a_x$ is the lattice constant along the $x$-direction, in which the resultant wavefunction is localized. The HWF can be thought of as a WF of a 1-dimensional system, coupled to the external parameters $k_y$ and $k_z$. It was proven that in one dimension exponentially localized WFs can always be found~\cite{Kohn-PR59, Kivelson-PRB82}. A nice generalization of this procedure to 3D is given in Ref.~\cite{Taherinejad-PRB14}.

\subsection{Wannier charge centers}
\label{ssec:wcc}
Given a set of Wannier functions, their charge centers are defined as the average position of charge of a Wannier function that resides in the home unit cell
\begin{equation}
\bar{\vec{r}}_n = \langle 0 n|\hat{\vec{r}}|0 n\rangle 
\end{equation}
Due to the ambiguity in the choice of the home unit cell, the Wannier charge centers (WCCs) are defined only modulo a lattice vector. Moreover, when the isolated group of bands in question contains more than one band, individual WCCs are not gauge-invariant~\cite{King-Smith-PRB93, Marzari-PRB97}. Only the sum of all WCCs is gauge-invariant modulo a lattice vector, and it is related to the electronic polarization~\cite{King-Smith-PRB93}. For a 1-dimensional system this relation reads
\begin{equation}
\vec{P}_{\rm e} =e \sum_n \bar{\vec{r}}_n
\label{eq:polar}
\end{equation}
where $e$ stands for the electronic charge. While $\vec{P}_{\rm e}$ is defined only up to a lattice vector, the {\it continuous} change $\Delta \vec{P}_{\rm e}$ under a continuous deformation of the Hamiltonian is a well-defined physical observable.   

A geometric interpretation in terms of the Zak phase~\cite{Zak-PRL89} can be given to WCCs. To do this, a Berry potential is introduced for the lattice periodic part of the Bloch functions as
\begin{equation}\label{eqn:berry_potential}
{\bf {\cal A}}_n({\bf k})=i\langle u_{n{\bf k}}|\bm{\nabla}_\vec{k}| u_{n{\bf k}}\rangle
\end{equation} 
In 1D WCCs can be redefined in terms of Berry potential using the transformations between Wannier and Bloch representations of Ref.~\cite{Blount-SSP62}
\begin{equation}
\bar{x}_n = \frac{i a_x}{2\pi}\int\limits_{-\pi/a_x}^{\pi/a_x} \mathrm{d}k_x~\langle u_{nk}|\partial_{k_x}|u_{nk}\rangle = \frac{a_x}{2\pi}\int\limits_{-\pi/a_x}^{\pi/a_x} \mathrm{d}k_x~\mathcal{A}_n(k_x)
\end{equation}
Similarly, the hybrid WCCs can be written as 
\begin{gather}
\bar{x}_n (k_y, k_z) = \bra{n; 0, k_y, k_z}\hat{r}_x\ket{n; 0, k_y, k_z} \\\nonumber
=  \frac{a_x}{2\pi} \int\limits_{-\pi/a_x}^{\pi/a_x} \mathrm{d}k_x~ \mathcal{A}(k_x, k_y, k_z).
\end{gather}
Thus, a hybrid WCC can be thought of as a WCC of a 1D system coupled to external parameters $(k_x, k_y)$. Since in crystalline systems $H({\bf k})=H({\bf k}+{\bf G})$, this coupling is equivalent to a periodic driving of a 1D system coupled to an external environment, as discussed in the context of charge pumping~\cite{Thouless-PRB83}. The existence of topological classification for such pumps was known long before the advent of topological materials~\cite{Thouless-PRB83, Thouless-PRL82}. We show below that many topological invariants of band structures can be obtained by studying the pumping of hybrid WCCs. A numerical procedure for constructing HWFs in a particular, ``maximally localized gauge,'' is given in Appendix~\ref{app:hwcc_computation}.
\subsection{Chern number via HWF}\label{ssec:chern_hwf}

The gauge field arising from the Berry potential (Eq.~\ref{eqn:berry_potential}) in a crystal is known as Berry curvature, and for a single isolated band it is defined as
\begin{equation}
\mathcal{F} = \nabla_\vec{k} \wedge \mathcal{A}(\vec{k}),
\end{equation}
where the wedge product in 3D is a usual cross product. 
In a multi-band case a non-Abelian Berry connection~\cite{Wilczek-PRL84} is introduced
\begin{equation}
\mathcal{A}_{mn, \alpha} = i \bra{u_{m\vec{k}}} \partial_\alpha \ket{u_{n\vec{k}}},
\end{equation}
and the corresponding gauge covariant formulation of Berry curvature in a multi-band case is
\begin{equation}
F_{mn, \gamma} = \mathcal{F}_{mn, \gamma} - \frac{i}{2} \varepsilon_{\alpha \beta \gamma} \left[\mathcal{A}_\alpha, \mathcal{A}_\beta \right]_{mn}.
\end{equation}
In 2D, or on a 2D cut of the 3D BZ, one can define a Chern number~\cite{Chern-AM46} of a single \emph{isolated} band as~\cite{Thouless-PRL82}
\begin{equation}
C_\gamma = \frac{1}{2\pi}\int_\text{BZ} \mathrm{d}^{2}k~\mathcal{F_\gamma}(\vec{k}),
\end{equation}
where $\gamma$ indicates the component normal to the 2D surface.
The corresponding equation for an isolated \emph{set} of bands is given by
\begin{equation}
C_{\gamma} = \frac{1}{2\pi}\int_\text{BZ}\mathrm{d}^2 k~\tr[F_{\gamma}]= \frac{1}{2\pi}\int_\text{BZ}\mathrm{d}^2 k~\tr[\mathcal{F_{\gamma}}],
\end{equation}
where the trace is taken over the band indices within the set. It can be shown~\cite{Soluyanov-thesis} that the same quantity can be written in terms of the hybrid WCCs
\begin{equation}
C = \frac{1}{a_x}\left(\sum\limits_n \bar{x}_n (k_y=2\pi) - \sum\limits_n \bar{x}_n (k_y=0)\right).
\label{cherncent}
\end{equation}
Here the WCCs $\bar{x}(k_y)$ are assumed to be smooth functions of $k_y$ for $k_y \in [0, 2\pi]$. This smoothness condition is fulfilled by constructing hybrid WCC in the 1D maximally localized gauge~\cite{Soluyanov-PRB11-b}. However, the periodicity condition is satisfied only modulo a lattice vector $R_x = n a_x$, where $n \in \Z$.

From this formulation one can see~\cite{Thouless-JPC84, Brouder-PRL07, Panati-AHP07, Panati-CMP13, Fiorenza-AHP16} how a non-zero Chern number becomes an obstruction for defining smooth and periodic Bloch states for the set of bands in question. In 1D maximally localized WFs can always be constructed~\cite{Kohn-PR59, Kivelson-PRB82}. In particular, the parallel transport procedure described in Appendix \ref{app:hwcc_computation} produces WFs that are the eigenstates of the projected position operator $\hat{x}=\hat{P}_{\mathrm{b}}\hat{X}\hat{P}_{\mathrm{b}}$, where $\hat{P}_{\mathrm{b}}$ is the projector onto the isolated bands \cite{Kivelson-PRB82}. In 2D, however, the projected position operators for the $x$ and $y$ coordinates in general do not commute~\cite{Marzari-PRB97}, and no set of WFs can be chosen to be maximally localized in both dimensions at the same time. Exponential localization, though, can still be achieved in both dimensions, unless the Chern number of the bands is non-zero~\cite{Marzari-PRB97, Brouder-PRL07}.  For a set of bands with a non-zero Chern number it is impossible to find a set of WFs exponentially localized in both dimensions: at least one WF is bound to have a power law decay in at least one direction in this case~\cite{Thouless-JPC84}. 

As mentioned above, the hybrid WFs are analogous to 1D WFs, but the 1D system here is coupled to external parameters (momenta in the other directions). For such a 1D system, the hybrid WF can still be chosen to be the eigenstate of the projected position operator, hence being exponentially localized in this direction. However, to analyze the charge pumping driven by the external parameters, continuity of the hybrid WCC in these parameters is required. If a Bloch band has a non-zero Chern number, the corresponding wave-function $\psi_{\bf k}$ cannot be chosen to be a smooth function of ${\bf k}$ in the interior of the BZ and still retain the periodicity condition $\psi_{{\bf k}+{\bf G}}=\psi_{\bf k}$, where ${\bf G}$ is a reciprocal lattice vector~\cite{Soluyanov-PRB12}. Thus, if one insists on a smooth evolution of the hybrid WCC $\bar{x}(k_y)$ as a function of $k_y$, the center of charge does not necessarily return to the initial position after a period of evolution $G_y$. However, crystalline periodicity guarantees that the center returns back to its position modulo a lattice vector, that is
\begin{equation}
\bar{x}(k_y)=\bar{x}(k_y+G_y) \mod R_x
\end{equation}  
Thus, Eq.~(\ref{cherncent}) illustrates how much charge is pumped through the 1D system during one continuous adiabatic cycle of the external parameter $k_y$.

This charge pumping is best understood as an externally induced change of electronic polarization of a 1D system in Eq.~\ref{eq:polar}. Using that expression of electronic polarization in terms of WCCs of a 1D system, and generalizing it to hybrid WCCs, the expression for the Chern number takes the form 
\begin{equation}
C=\frac{1}{ea}\left(P^{h}_{\mathrm{e}}(2\pi)-P^{h}_{\mathrm{e}}(0)\right),
\label{eqn:Chern}
\end{equation}
where we introduced hybrid electronic polarization as 
\begin{equation}
P^h_{\mathrm{e}} (k_y)= e \sum_n \bar{x}_n (k_y).
\end{equation}
According to the above, the value of HWCC are not gauge invariant. However, the hybrid polarization, and thus the Chern number, are gauge invariant. The Chern number reflects the obstruction for the possibility to construct maximally localized HWFs. Thus, Eq.~\ref{eqn:Chern} can give a correct Chern number even without maximally localized HWFs - other HWFs would work as well.

\begin{figure} %\centering
\includegraphics[width=\columnwidth]{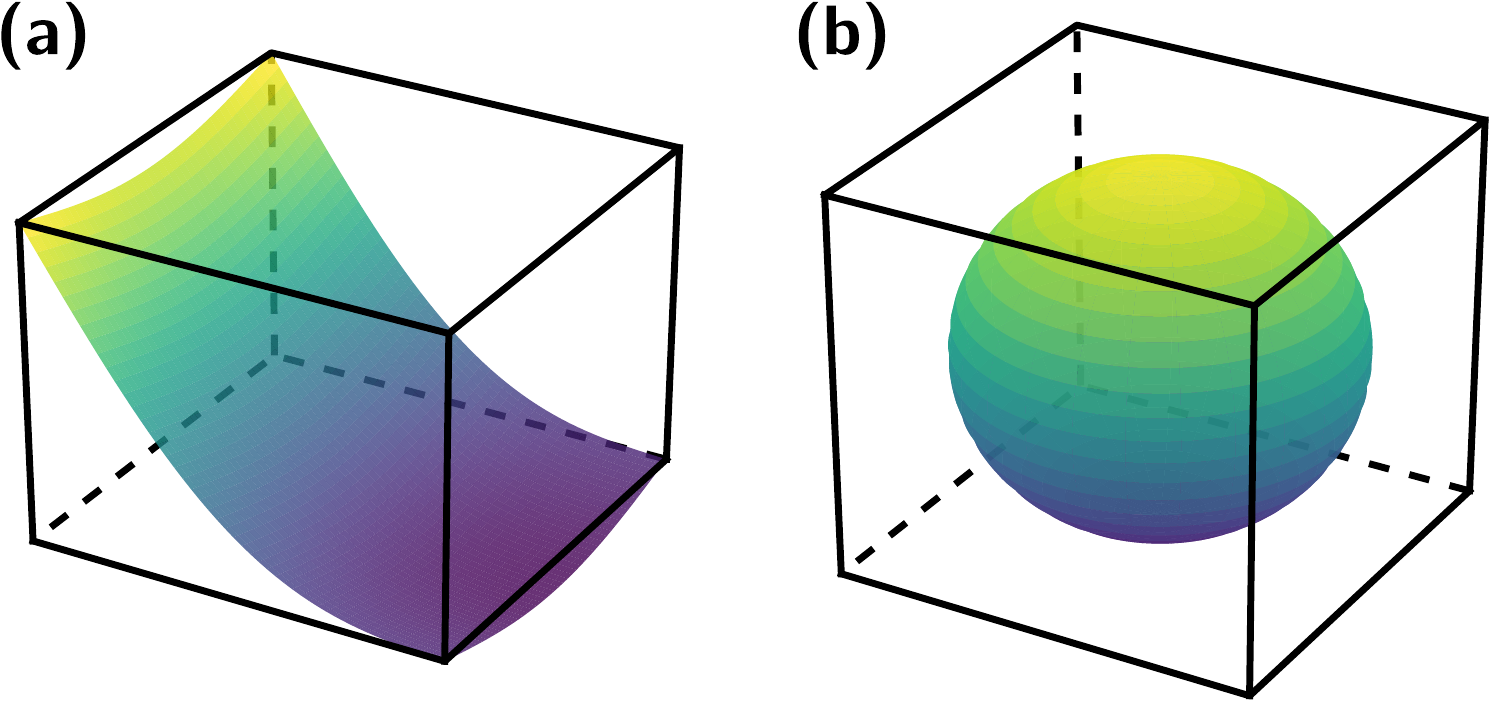}\qquad
\caption[]{Two examples (shaded) of possible closed 2D cuts in a cubic Brillouin zone. Since periodicity is imposed at the Brillouin zone boundary, the surface in panel (a) is topologically equivalent to a torus.}
\label{fig:surface_curved}
\end{figure}
In 3D, the Chern number can be defined for any closed 2D cut of the BZ. Possible examples of such cuts are shown in Fig.~\ref{fig:surface_curved}.

\begin{figure}[h]\centering
\parbox[t]{\columnwidth}{
\includegraphics[width=\columnwidth]{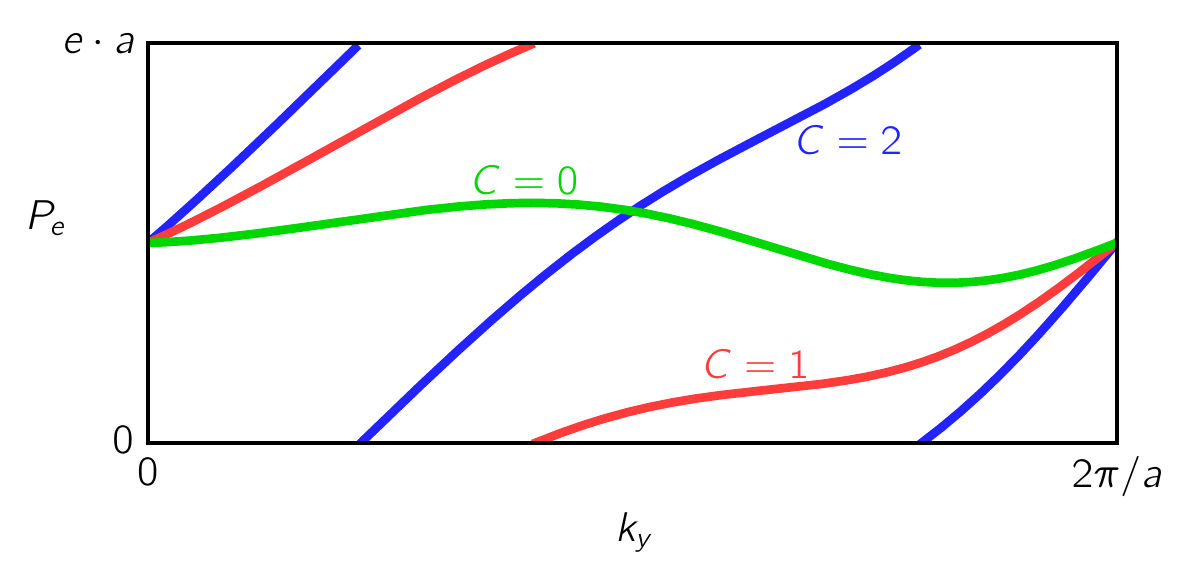}
\includegraphics[width=0.85\columnwidth]{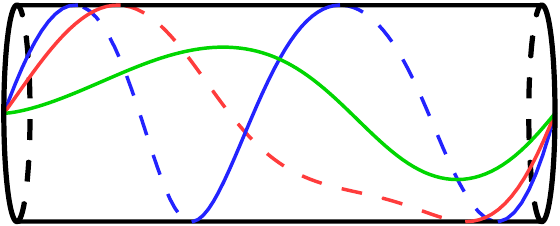}
\caption[]{Sketch of some possible evolutions of polarization ($P_e(k_y)$) across the BZ, exhibiting different Chern numbers $C$.}
\label{fig:chern}}
\end{figure}

According to the discussion above the hybrid WCC $\bar{x}(k_y)$ is defined only modulo a lattice vector, reflecting the periodicity of the lattice in real space. It is thus convenient to put periodic boundary conditions on the hybrid WCC of the form $\bar{x}(k_y)=\bar{x}(k_y)+R_x$, so that $\bar{x}$ represents a point on a circle $S^1$ at each $k_y$. This point gives the position of the center of charge in the unit cell for a given value of $k_y$. Given that the WCCs are assumed to be smooth functions of $k_y\in [0, 2\pi]$, the pump can be visualized as the flow of WCCs on the surface of the cylinder $S^1\times [0,2\pi]$, as shown in Fig.~\ref{fig:chern}. Since the hybrid electronic polarization is a sum of all hybrid WCCs, it is also defined on the surface of this cylinder. The Chern number is then associated with the number of windings $P_{\mathrm{e}}^h(k_y)$ performs around the cylinder when $k_y$ is changed from $0$ to $2\pi$~\cite{Soluyanov-PRB11-b, Soluyanov-thesis}. 
Thus, the Chern number can be associated with the number of unit cells traversed by the net center of charge of all bands within the pumping cycle, which can be equivalently thought of as the number of electronic charges pumped across one unit cell in the course of a cycle. This interpretation makes the relation between the Chern number and the Hall conductance explicit.

%
%If the electronic polarization is defined in such a way that it evolves smoothly on the interval $k_y \in [0, 2\pi]$, the Chern number can be defined as the change in polarization across this interval:
%
%\begin{equation}
%C = P_e(2\pi) - P_e(0)
%\end{equation}
%
%This invariant is protected by the fundamental properties of a crystalline insulator, translational symmetry and charge conservation.

%
\subsection{Wilson loops and gauge choices}
An alternative physical understanding of the hybrid WCCs was proposed in Refs.~\cite{Yu-PRB11, Soluyanov-PRB11-a}, for the special case of TR-symmetric systems. It was shown that the eigenvalues of the projected position operator represents the Wilson loop of the $U(2N)$ non-Abelian Berry connection, where $2N$ is the number of occupied states in a TR-symmetric insulator~\footnote{This number has to be even as a consequence of Kramers theorem.}. This Wilson loop approach was later generalized to other topological phases~\cite{Alexandradinata-PRB14, Alexandradinata-PRL14, Alexandradinata-PRB16, Taherinejad-PRB14}. Here we discuss the differences between hybrid WCC and the Wilson loop approaches.

The Wilson loop for the non-interacting insulating systems is defined~\cite{Yu-PRB11} in terms of the projector onto the $N_{\rm{occ}}$ occupied states as
\begin{equation}
P^{\rm{occ}}_{\bf k}=\sum_{j=1}^{N_{\rm{occ}}} | u_{j{\bf k}}\rangle\langle u_{j{\bf k}}|.
\label{eqn:proj_occ}
\end{equation}
Given a closed curve ${\cal C}$ in $k$-space, discretized in $L$ points $k_i$, $i=0,..,L-1$, the Wilson loop is computed as
\begin{equation}
W({\cal C})=\prod_{i=0}^{L-1} P^{\rm{occ}}_{{\bf k}_i}
\label{eqn:wilson}
\end{equation} 
and is a $N_{\rm{occ}}\times N_{\rm{occ}}$ matrix.
%(see Appendix~? for the explanation of this formula and its applicability). 
Wilson loops are known to be a gauge invariant quantity in quantum field theory~\cite{Peskin-book}, and this is still the case in the present definition. Indeed, the gauge transformation of the form~\ref{eqn:gauge} leaves $W({\cal C})$ invariant. By taking the $\log$ of the eigenvalues of the Wilson loop at an arbitrary point on the loop $\cal C$ (and normalized by $2\pi$), one arrives at a special gauge-invariant set of hybrid WCCs, which exactly coincide with those obtained from the maximally localized WCC construction outlined in App.~\ref{app:hwcc_computation}.

From the theory of polarization~\cite{King-Smith-PRB93} reviewed above it is known that the WCCs are in general gauge dependent. To reconcile this with the gauge-independence of the Wilson loop, notice that the projector onto the occupied space can be equivalently considered to be a sum of projectors onto different subspaces comprising the occupied space
\begin{equation}
P^{\rm{occ}}_{{\bf k}}=\sum_{\ell=1}^D P^{(\ell)}_{\bf k},
\label{proj1}
\end{equation}
where $D$ is the number of subspaces. Various examples of such splittings are discussed below, but the simplest of them is given by a large set of isolated bands, that is composed of $D$ smaller isolated sets. Then each of the sets can be treated separately, and each of the projectors $P^{(\ell)}_{\bf k}$ can be used separately to construct a Wilson loop $W_\ell ({\cal C})$ for each of the isolated sets of bands separately.

The hybrid WCC obtained by diagonalizing the Wilson loops $W_\ell$ are in general different from the ones obtained by diagonalizing the full loop $W$. This can be seen by noticing that when constructing $W$ with the projector~(Eq. \ref{proj1}), cross terms of the form $P^{(\ell)}_{{\bf k}_i}P^{(\ell^\prime)}_{{\bf k}_{i+1}}$ ($\ell \neq \ell^\prime$) will appear, while they are absent when constructing $W_\ell$'s separately. Note, that in accordance with the theory of polarization, both constructions are physically equivalent, since the sum of the hybrid WCCs at each point on the curve ${\cal C}$ will be the same for the two constructions (modulo a quantum), corresponding to 1D electronic polarization.

At the risk of abuse of terminology we refer to the various ways of constructing the WCCs as a gauge freedom. This is motivated by the definition of the WFs and hybrid WFs -- their construction depends on a particular choice of representative Bloch states $\psi_{n{\bf k}}$ used to represent the Hilbert space of interest.   

\section{General strategy for identifying topological materials}
\label{sec:strat}

While the net Chern number is protected by charge conservation, less fundamental symmetries can exist and also induce a topological classification. These topological classes are in general not captured by the net Chern number. In this section, we propose a general route to robust identification of such topological states in both real materials and models based on the notion of individual Chern numbers~\cite{Soluyanov-PRB12}. We first explain what individual Chern numbers are, and then show how they can be used to track down the presence of symmetry-protected topological order in materials and calculate the corresponding topological invariants. One of the clear examples of such a procedure are the mirror Chern numbers~\cite{Teo-PRB08}.

\subsection{Individual Chern numbers}\label{ssec:individual_chern}
The notion of individual Chern numbers \cite{Soluyanov-PRB12} is based on the idea of splitting the Hilbert space spanned by an isolated set of bands ${\cal H}_\mathrm{set}$ into a collection of Hilbert spaces
\begin{equation}
{\cal H}_\mathrm{set} = \bigoplus_{i=1}^N {\cal H}_i
\end{equation}
in such a way that the Chern number associated with each of these Hilbert spaces is an integer. This means that the projector $P_\vec{k}^\mathrm{set}$ onto ${\cal H}_\mathrm{set}$ is decomposed into projectors on the individual Hilbert spaces
\begin{equation}
P^{\rm{set}}_{\bf k}=\sum_{i=1}^{N} P^{(i)}_{\bf k}
\end{equation}
for any $\vec{k}$ on the 2D smooth and closed manifold $M$, on which the bands are defined. The necessary condition for the individual Chern numbers to be integral is that each projector $P_\vec{k}^{(i)}$ is smooth on $M$~\cite{Prodan-NJP10}.

The total Chern number of the set of bands~\cite{Prodan-NJP10}
\begin{equation}\label{eqn:chern_projector}
C_\mathrm{set} = \frac{i}{2\pi} \int\limits_M \tr \left\{P_\vec{k}^\mathrm{set} \left[\partial_{k_1} P_\vec{k}^\mathrm{set} , \partial_{k_2} P_\vec{k}^\mathrm{set} \right] \right\} \mathrm{d}k_1 \wedge \mathrm{d}k_2
\end{equation}
is then equal to the sum of individual Chern numbers (see Appendix~\ref{app:individual_chern})
\begin{equation}
C_\text{set}=\sum_{i=1}^N c_i,
\label{ctot}
\end{equation}
where
\begin{equation}
c_i = \frac{i}{2\pi} \int\limits_M \tr \left\{P_\vec{k}^{(i)} \left[\partial_{k_1} P_\vec{k}^{(i)} , \partial_{k_2} P_\vec{k}^{(i)} \right] \right\} \mathrm{d}k_1 \wedge \mathrm{d}k_2.
\end{equation}

\begin{figure}\centering
\includegraphics[height=7.2cm]{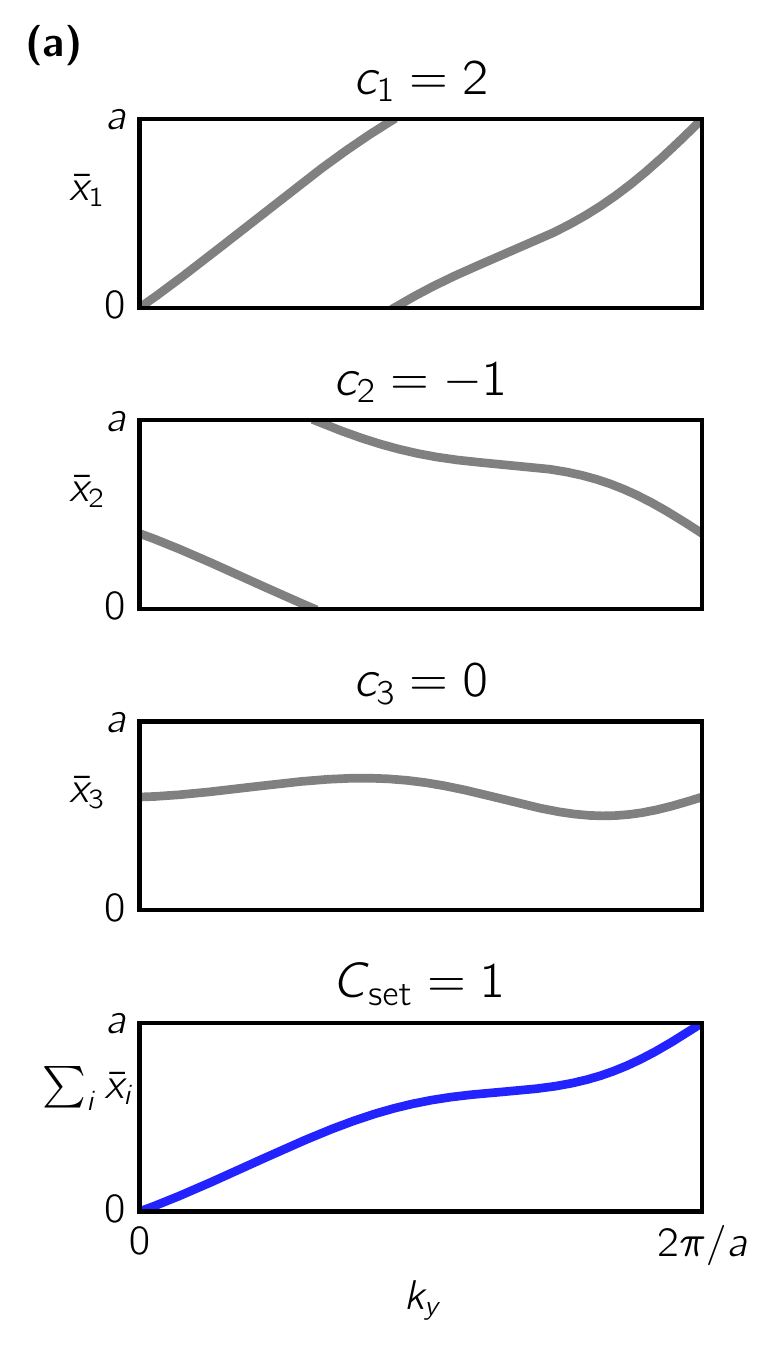}\quad
\includegraphics[height=7.2cm]{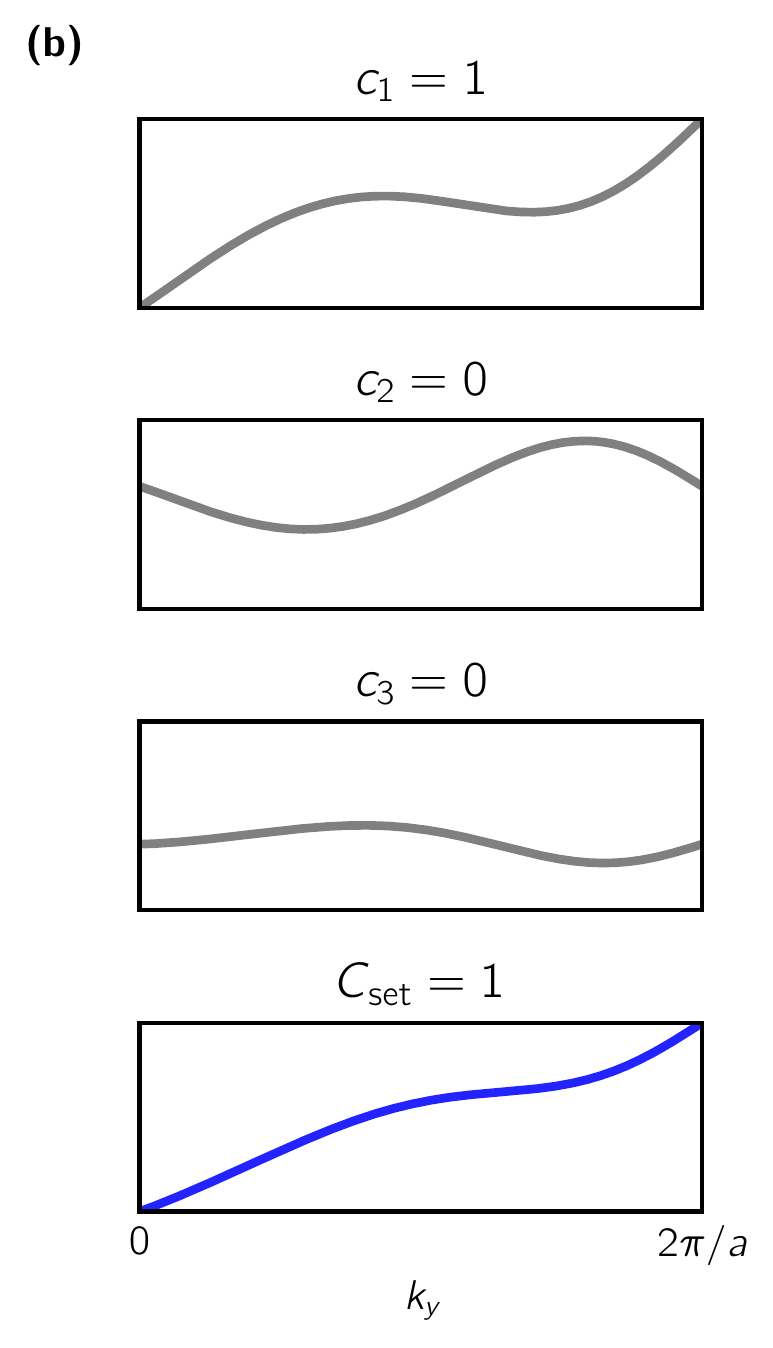}
\caption[]{Illustration of different gauge choices for a system with $N=3$ occupied bands and $C_\text{set}=1$. Left panel: Bands No. $1$ and $2$ have individual Chern numbers $c_1=2$ and $c_2=-1$. Right panel: Bands No. $2$ and $3$ have zero Chern number, and thus $c_1 = C_\text{set} = 1$}
\label{fig:individual_chern}
\end{figure}

An important example of such splittings of the Hilbert space is the one with each ${\cal H}_i$ containing only a single band $\ket{u_{\vec{k}, i}}$. If the projectors $\ketbra{u_{\vec{k}, i}}{u_{\vec{k}, i}}$ are  chosen to be smooth on the manifold $M$, each band is assigned an integer individual Chern number. %Since such projectors are always real, a phase winding of the Bloch bands does not necessarily break the smoothness of the projectors. 
However, an actual construction of a gauge (projector choice) that results in integer individual Chern numbers is a complicated task~\footnote{See Ref.~\cite{Soluyanov-PRB12} for an example of such a construction for quantum spin Hall systems, and Ref.~\cite{Winkler-PRB16} for a general construction in 2 and 3 dimensions.}, since the gauge (and hence projectors) obtained from diagonalization of the Hamiltonian numerically on the mesh of $k$-points can have discontinuities around degeneracy points in the energy spectrum. Moreover, in accord with the above discussions of WCCs and Wilson loops, one can manipulate the gauge choice to produce a different decomposition of ${\cal H}_\mathrm{set}$ with different values of $c_i$'s, as illustrated in Fig.~\ref{fig:individual_chern}.

Since the individual Chern numbers depend on a particular splitting of the Hilbert space, they lack physical meaning, unless some physical constraints on the subspaces ${\cal H}_i$ fix their values. These constraints are provided by the symmetries of the underlying Hamiltonian. If the gauge used to split the Hilbert space into individual Bloch states respects the symmetry (meaning that the projectors resulting from the splitting respect the symmetry), a symmetry-protected topological phase could have at least some non-zero individual Chern numbers. Thus, choosing the subspaces according to their symmetry behavior could allow for a classification of symmetry-protected topological states. Whether this classification is unique or complete is, at present time, unknown. 

An illustrative example here is provided by time-reversal (TR) invariant systems~\cite{Kane-PRL05-a, Kane-PRL05-b}, where Bloch bands come in Kramers pairs. Consider the case of a single such pair. An individual Chern number can be associated with each of the two bands in the Kramers pair. This is equivalent to splitting the Hilbert space spanned by the Kramers pair into two subspaces 
\begin{equation}
P_{\bf k}=P^{(1)}_{\bf k}+P_{\bf k}^{(2)}.
\end{equation}  
corresponding to projectors $P_\vec{k}^{(i)}$ that are smooth on $M$. In the gauge that respects TR symmetry, the two projectors are related by
\begin{equation}
P^{(1)}_{\bf k}=\theta P_{-{\bf k}}^{(2)}\theta^{-1},
\label{TRconstr}
\end{equation}
for all $\vec{k}$ in $M$, where $\theta$ is the TR operator. Under this constraint, the HWCCs need to come in pairs of TR-symmetric momenta~\cite{Fu-PRB06, Soluyanov-PRB11-b}, and the individual Chern numbers of the two bands must be opposite ($c_2 = -c_1$). In the quantum spin Hall phase, they are constrained to be odd, while they are bound to be even in the $\mathbb{Z}_2$-even phase~\cite{Fu-PRB06}, as illustrated in Fig.~\ref{fig:z2}. It is TR-symmetry that enforces the distinction between the two phases, and no splitting of the Kramers pair in the $\mathbb{Z}_2$-odd phase subject to the TR constraint of Eq.~\ref{TRconstr}, can produce vanishing individual Chern numbers, thus proving the robustness of topological phase protected by TR-symmetry. Due to the particular symmetry of the HWCCs, however, it is possible to distinguish the $\mathbb{Z}_2$-even and -odd phases even without explicitly calculating the individual Chern numbers~\cite{Fu-PRB06,Soluyanov-PRB11-a, Yu-PRB11}, as will be illustrated in Sec.~\ref{ssec:z2} for a realistic many-band case.

More illustrations of symmetry-protected individual Chern numbers are provided below. The general approach is to construct HWCCs on certain surfaces in the BZ in a gauge that respects a symmetry of the Hamiltonian, to see whether this symmetry protects non-zero individual Chern numbers. As shown below this approach can be readily used to identify the known topological phases of non-interacting systems. 

\subsection{Application to the search for topological materials}
Based on the properties of individual Chern numbers, we outline several cases of this procedure that can potentially predict topological materials. As previously discussed, for a symmetry-protected topological phase, the gauge that respects the symmetry protecting the topology results, in all the cases studied so far, in non-zero individual Chern numbers. This implies a gapless flow of HWCCs on some symmetry-respecting surfaces in the BZ. While a robust confirmation of the possible presence of the gapless HWCC flow (and hence, a topological phase) might require additional analysis in some cases,  its absence is often easy to see. Consequently, possible candidate materials can be identified by screening the high-symmetry planes in the BZ for the presence of a gapless flow of the WCCs constructed in the symmetry-preserving gauge.

Once a candidate material is identified, the next step is to uniquely define its topology. In some (but not all) cases this is simply a matter of calculating the total Chern number or the $\mathbb{Z}_2$ invariant on specific surfaces in reciprocal space. In a more general case, this is done by splitting the Hilbert space into subspaces according to their symmetry behaviour, as described in Appendix \ref{app:symmetry_op}.
\section{Application to insulators}
\label{sec:insulators}

In this section, we discuss and illustrate how the  Wannier centers flow  is applied to insulators. The discussion covers the cases of Chern (quantum anomalous Hall) insulators,  TR-symmetric $\mathbb{Z}_2$ topological insulators and crystalline topological insulators, including those, where topology is protected by rotational symmetries.

\subsection{Chern insulator}
\label{ssec:chern}

Chern insulators are 2D materials with broken TR-symmetry, in which the occupied Bloch bands have a non-zero total Chern number, which is the topological invariant characterizing this phase~\cite{Haldane-PRL88, Volovik-88, Thouless-PRL82}. The Chern number takes on integer values, and these values correspond to the integer Hall conductance in units of $2e^2/h$ exhibited by the material in the absence of an external magnetic field. Due to the presence of robust chiral edge states, these materials are expected to be useful in many technological applications %~\cite{?}
. Several compounds were predicted to host this phase~\cite{Liu-PRL08, Qiao-PRB10, Yu-Sci10, Kou-JAP12, Xu-NatPhys12, Niu-APL11, Jiang-PRB12, Zhang-PRB13, Zhang-ARX11, Garrity-PRL13}, and experimental evidence of its existence in some of them was found experimentally~\cite{Chang-Sci13}. However, no stoichiometric crystalline material was experimentally identified yet, and the quest for a wide-gap Chern insulator material is still ongoing at the time of writing.

The search strategy, supplemented by Z2Pack, has what we hope will be two promising directions.  One is to simulate thin films of magnetic materials directly and compute the Chern number of such effectively 2D systems. The total hybrid polarization is tracked and plotted as a function of momentum, giving the value for the topological invariant via the hybrid Wannier function approach described above in Sec.~\ref{ssec:chern_hwf}. Such a route to search for Chern insulators should also be taken when simulating heterostructures, such as quantum wells, or interfaces that are candidate for this phase. Another candidate platform for realizing a Chern insulator is provided by thin films of magnetic (semi)-metals, where quantum confinement due to finite size can lead to a bulk gap opening, making the thin film insulating. Such simulations require the use of supercells, so the direct implementation can be computationally expensive when a realistic description of the system requires the use of methods beyond the standard density functional theory, such as hybrid functionals~\cite{Hedin-SSP70, Becke-JCP93, PBE0, Heyd-JChPh03} or $GW$~\cite{GW}. In such cases one can use Z2Pack to identify the Chern numbers for the corresponding tight-binding models.

A different approach, especially suitable for the search for a stoichiometric crystalline Chern insulator, does not require the use of supercells and consists in finding a 3D material with a 3D quantum Hall effect - which is necessarily a layered compound of 2D Chern insulators. It is motivated by the observation, that in an insulating material 2D cuts of the 3D BZ represent a BZ of some imaginary 2D insulator. Thus, Chern numbers can be defined on different 2D cuts in the BZ, giving a classification of magnetic materials in terms of a set of 3 Chern numbers~\cite{Halperin-JJAP87}.

\begin{figure}\centering
\includegraphics[width=0.8\columnwidth]{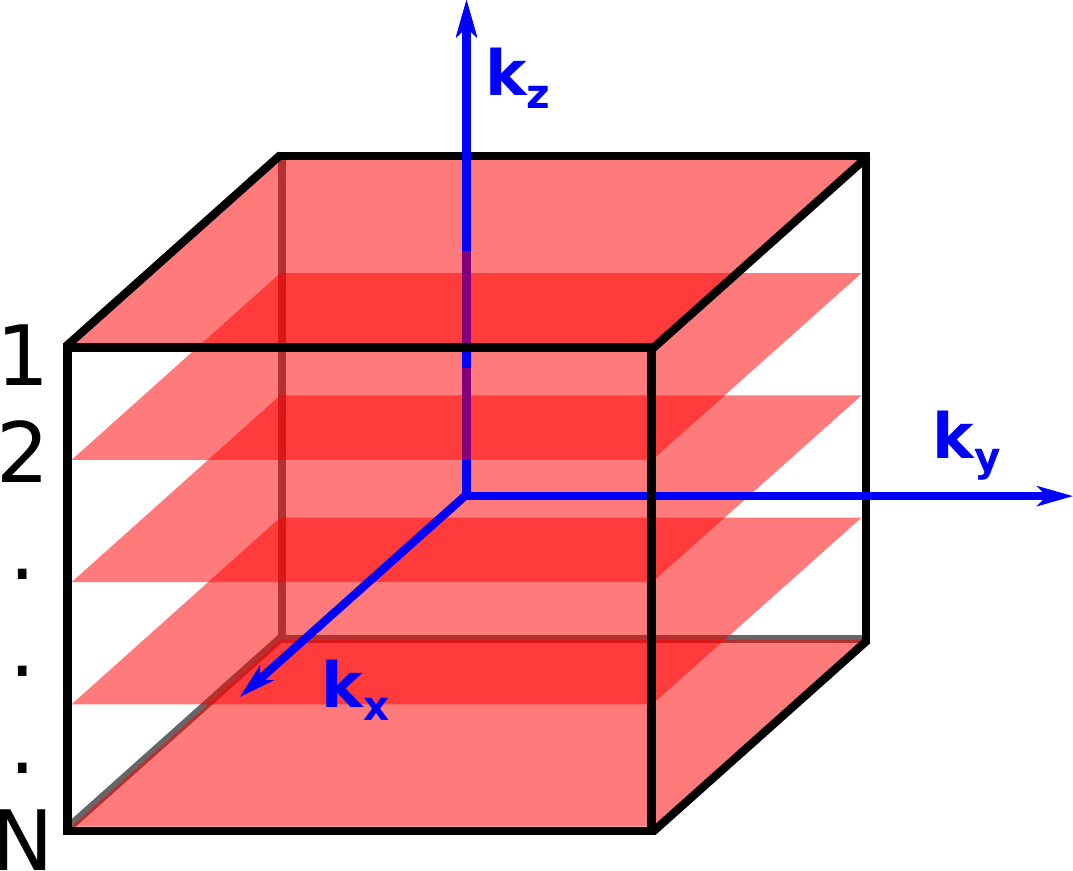}
\caption[]{Sketch of a cubic Brillouin zone. In an insulator, the Chern numbers $C_1, ..., C_N$ associated with surfaces $1, ..., N$ orthogonal to the $k_z$ direction (shown in red) are all equal. The same is true for surfaces perpendicular to $k_x$ or $k_y$.}
\label{fig:cubic_bz}
\end{figure}

For example, consider a cubic magnetic insulator with a cubic BZ as shown in Fig.~\ref{fig:cubic_bz}. Taking 2D square cuts of the BZ at fixed values of $k_i$, $i=x,y,z$ allows to define a 2D Chern number on any of these cuts. Moreover, one can argue that the cuts taken at $k_i$ and $k_i+\delta k_i$, have the same Chern number, since in going from the 2D cut at $k_i$ to the one at $k_i+\delta k_i$, the band gap does not close on the 2D cut, and thus the 2D systems at these two momenta represent 2D systems that can be adiabatically connected without closing the band gap. Thus the Chern number of all the 2D cuts taken for a certain $k_i$ have to be the same.

More generally, the Chern number is invariant under an adiabatic change in the manifold. Also, the Chern number of a union of disjoint manifolds is the sum of their Chern numbers. Using these two simple rules, the Chern number of other closed manifolds can be inferred from the three described above. For example, the plane defined by $k_x=k_y$ can be smoothly transformed into a sum of the two planes at $k_x=0$ and $k_y=2\pi / a$. Once a BZ cut with a non-zero Chern number is identified, the surfaces of the crystal that exhibit the quantum anomalous Hall effect are known. 

While these two methods so far represent wishful thinking in terms of finding materials, we hope that a thorough search of databases based on these methods and using our code to compute Chern numbers will lead to the discovery of the first Chern insulator. 

%\textcolor{red}{Add a sentence or two on how the Chern numbers of other planes can be inferred from these three {\color{blue}{(done)}}, and that the corresponding surfaces, or thin films will exhibit the quantum anomalous Hall effect.} {\color{blue}{Isn't that what's written in the intro to this section?}}

\subsection{$\mathbb{Z}_2$ phases}\label{ssec:z2}
Here we discuss the numerical determination of the $\mathbb{Z}_2$ topological phases in materials. The numerical method for computing this topological invariant was first introduced in Ref.~\cite{Soluyanov-PRB11-b} in the context of TR-symmetric topological insulators~\cite{Kane-PRL05-b}. However, it can be equally applicable to any system, in which the Hilbert space of interest can be split into two symmetry-related subspaces, each of which has an odd individual Chern number. We summarize the method of Ref.~\cite{Soluyanov-PRB11-b} below, and put it in the context of individual Chern numbers.

\subsubsection{$\mathbb{Z}_2$ classification due to TR symmetry}
For 2D systems, the $\mathbb{Z}_2$ classification distinguishes two topological phases. There are only two classes of gapped Hamiltonians that cannot be adiabatically connected without closing the band gap or breaking the classifying symmetry.

The standard example here is given by TR-symmetric topological insulators. These are classified into two distinct classes~\cite{Kane-PRL05-b} depending on the number of Kramers pairs of edge states appearing at a semi-infinite 1D edge of such an insulator. By changing the Hamiltonian adiabatically while preserving TR symmetry, these Kramers pairs can only be removed from the edge spectrum {\it in pairs}. Thus, Hamiltonians hosting an odd number of Kramers pairs of edge states at the boundary are topologically distinct from those that host an even number of such pairs~\cite{Kane-PRL05-b}. 

The bulk bands in these insulators come in Kramers pairs of states related by TR-symmetry. The case of a single occupied Kramers pair was briefly mentioned above in Sec.~\ref{ssec:individual_chern}. For an arbitrary number of Kramers pairs, the occupied Hilbert space ${\cal H}_\mathrm{set}$ is split into two subspaces ${\cal H}_{1,2}$, such that the projectors onto each of them are smooth and related by TR-symmetry, as in Eq.~\ref{TRconstr}. Now, however, both $P_1$ and $P_2$ are projectors onto a set of bands, rather than just a single band. In TR-symmetric systems the net Chern number has to vanish, since it is odd under TR. Thus, the two subspaces related by TR-symmetry necessarily have opposite individual Chern numbers.

When choosing $\mathcal{H}_{1, 2}$, each Kramers pair is split into two states that are assigned to different subspaces. As long as the two states are mapped onto each other by TR-symmetry, it does not matter how this assignment of states to the subspaces is done. Indeed, the states can exchange subspaces. Since the states carry opposite Chern numbers, this exchange can only change the individual Chern numbers $C_{1, 2}$ of the subspaces by an even number. Thus, a $\mathbb{Z}_2$ invariant can be defined as
\begin{equation}
\Delta = (C_{1}-C_{2})/2 \mod 2.
\end{equation}

For the topological quantum spin Hall phase, $C_{1,2}$ are odd, so that TR-symmetry does not allow for the construction of smooth Bloch states spanning both these subspaces~\cite{Loring-EPL10, Soluyanov-PRB11-a}. Since the Chern numbers of the subspaces represent the change of their corresponding electronic polarizations, the invariant $\Delta$ can be defined via the notion of TR-polarization, defined in Ref.~\cite{Fu-PRB06}.

The $\mathbb{Z}_2$ phases of 3D materials are classified by a set of indices~\cite{Fu-PRL07, Moore-PRB07, Roy-PRB09-a, Roy-PRB09-b}
\begin{equation}
\nu; (\nu_x, \nu_y, \nu_z),
\end{equation}
defined through the 2D invariants on the TR-invariant planes in the BZ
\begin{gather}
\nu = \Delta(k_i = 0) + \Delta(k_i = 0.5) \mod 2\\
\nu_i = \Delta(k_i = 0.5),
\end{gather}
where $k_i$ is in reduced coordinates. A system is called a weak topological insulator if any of the $\nu_i$ is non-zero but $\nu=0$, while a system with $\nu=1$ is referred to as a strong topological insulator~\cite{Fu-PRL07}.

Note, that the definition is in terms of invariants $\Delta$ of those cuts of the 3D BZ that can be considered as BZs of some 2D TR-symmetric insulators. A 2D $\mathbb{Z}_2$ invariant $\Delta$ can be defined on any plane in the BZ that for each point $\vec{k}$ also contains its TR-image $-\vec{k}$.

\subsubsection{$\mathbb{Z}_2$ phase in terms of hybrid Wannier functions}

In practice, splitting the two occupied space of a TR-symmetric insulator into two subspaces ${\cal H}_{1,2}$ related by TR, and spanned by smooth projectors is a non-trivial task. While in the presence of additional symmetries~\cite{Sheng-PRL06, Prodan-PRB09} such a splitting is possible, it is preferable to have a numerical method for computing $\Delta$, which does not require an explicit splitting of the Hilbert space. Such a formulation is given in terms of HWCCs~\cite{Soluyanov-PRB11-b}, and we recap it here. 

\begin{figure}[h]\centering
\includegraphics[width=\columnwidth]{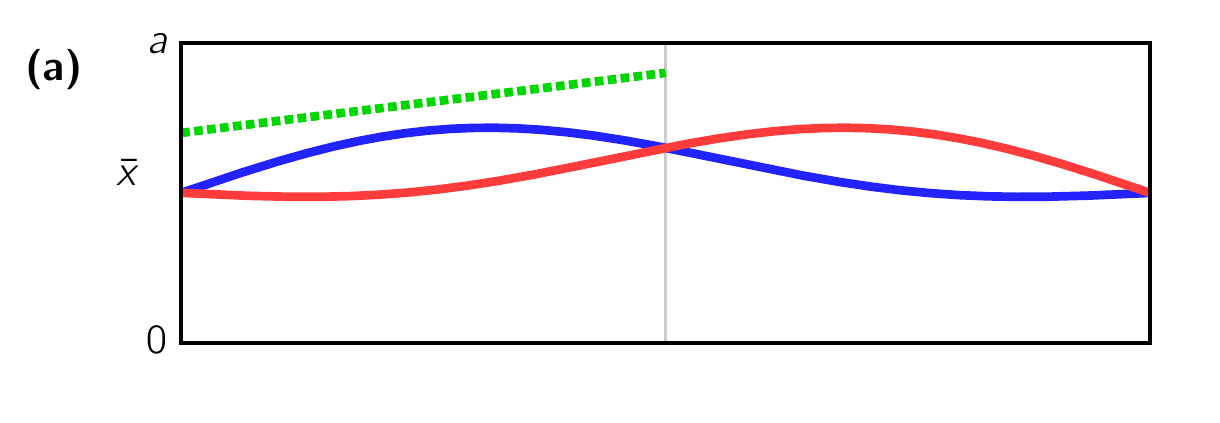}
\includegraphics[width=\columnwidth]{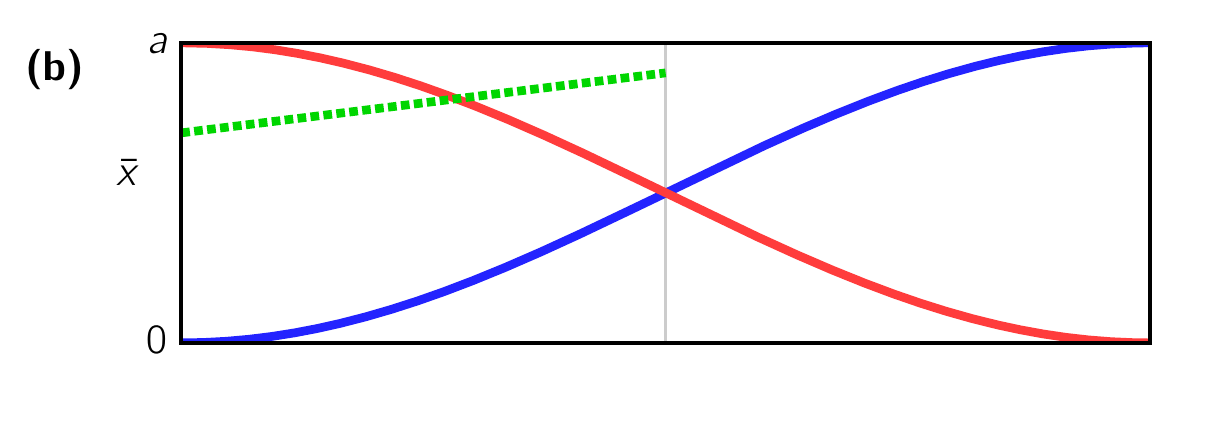}
\includegraphics[width=\columnwidth]{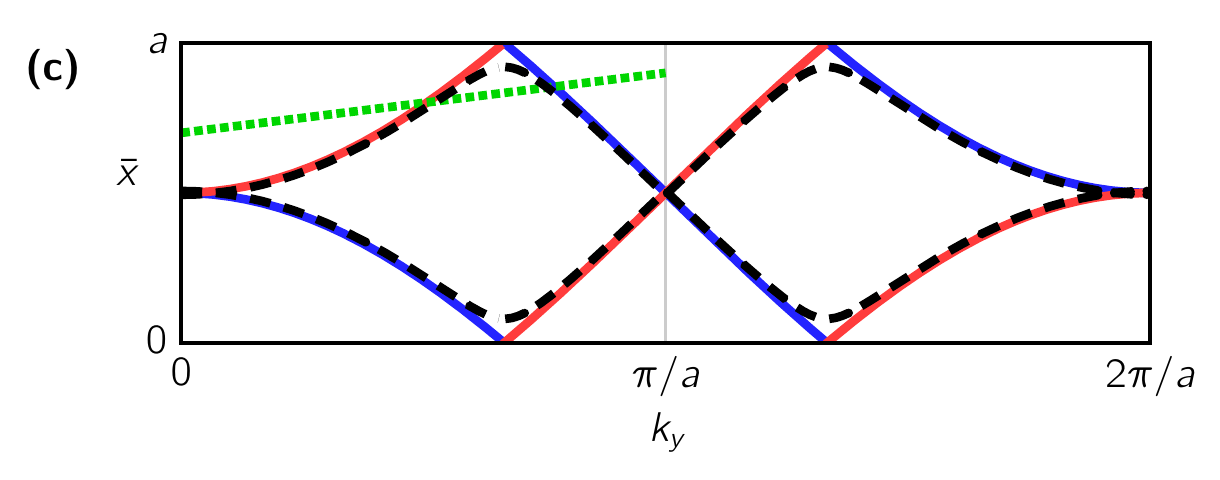}
\caption[]{Different possible WCC evolutions (red and blue lines) for a system with 2 occupied bands and time-reversal symmetry. The $\mathbb{Z}_2$ invariant can be calculated from the number of WCC crossings $L$ of an arbitrary line $x_\text{cut}(k_y)$ (dotted green line) across half the BZ. (a) Both WCC have winding number $0$, corresponding to a $\mathbb{Z}_2$ trivial state ($L = 0, \Delta = 0$). (b) WCC with winding numbers $\pm 1$, corresponding to a $\mathbb{Z}_2$ non-trivial state ($L = 1, \Delta = 1$). (c) WCC with winding numbers $\pm 2$. Because the crossings at momenta other than $k_y=0, \pi$ are not protected, the system is adiabatically connectable (dashed black lines) to the one in (a) ($L=2, \Delta = 0$).}
\label{fig:z2}
\end{figure}

In a gauge that respects TR-symmetry, the HWCC come in pairs
\begin{equation}
\bar{x}_{2j - 1}(k_y) = \bar{x}_{2j}(-k_y) \mod a_x, 
\label{eqn:kramers_pair}
\end{equation}
for any given $k_y$. Consequently, they are equal up to a lattice constant at the special points $k_y = 0$, $\pi / a_y$ and $2\pi / a_y$. This condition allows for two distinct topological phases, illustrated in Fig.~\ref{fig:z2} for a single Kramers pair. 

\begin{figure}\centering
\includegraphics[height=3.95cm]{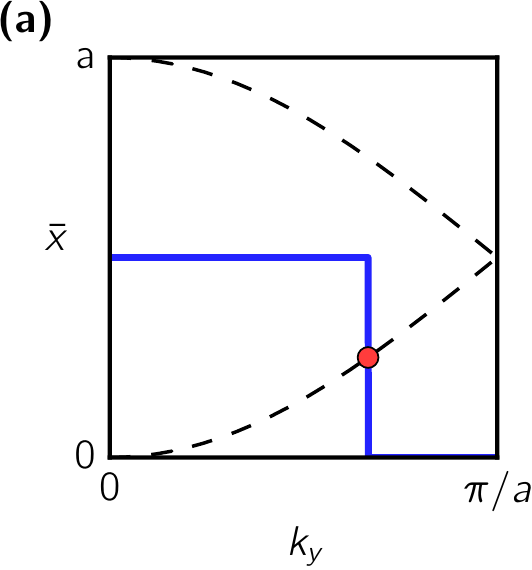}\quad
\includegraphics[height=3.95cm]{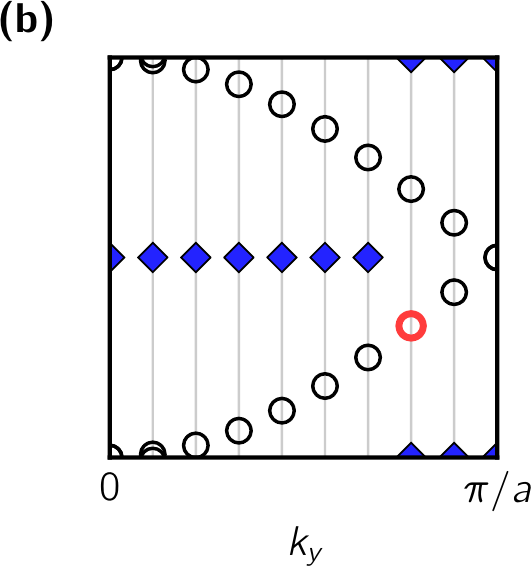}
\caption[]{Sketch of a $\mathbb{Z}_2$ calculation. (a) Continuous illustration of HWCCs. The largest gap (blue line) between any two HWCC (dashed lines) crosses a single HWCC (red dot). (b) Discrete illustration of HWCCs. The crossing between HWCC (circles) and the largest gap (blue rhombi) is found by searching for a HWCC lying between neighboring gaps.}
\label{fig:cut}
\end{figure}

The $\mathbb{Z}_2$ invariant is computed by considering the HWCCs $\bar{x}(k_y)$ for only half of the momentum values (that is $k_y \in [0,\pi / a_y]$), since the other half is symmetric in the TR-respecting gauge (for each HWCC at $k_y$ there exists its TR-image at $-k_y$). The invariant is given by the number of times $L$ any line $x_\mathrm{cut}(k_y)$ crosses a HWCC line when going from $k_y=0$ to $k_y=\pi / a_y$~\cite{Soluyanov-PRB11-b, Yu-PRB11} (see Fig.~\ref{fig:z2}). This number will be even if the system is in a $\mathbb{Z}_2$-trivial state and odd otherwise. Thus, the $\mathbb{Z}_2$ invariant is simply
\begin{equation}
\Delta = L \mod 2.
\end{equation}
For reasons of numerical convergence and to avoid plotting the HWCCs, it is best to define $x_\mathrm{cut}(k_y)$ as the largest gap between any two HWCCs at a given $k_y$~\cite{Soluyanov-PRB11-b}. The situation is illustrated in the Fig.~\ref{fig:cut} for a single Kramers pair, where a color scheme shows the possible splitting of the occupied Kramers pair into two TR-related states with opposite individual Chern numbers.
On a discrete mesh, $L$ is computed by counting the number of HWCC lying between neighboring values of the largest gap $x_\text{cut}$~\cite{Soluyanov-PRB11-b}. The details of the numerical calculation are described in Appendix~\ref{app:z2_computation}.

For completeness, we illustrate the use of Z2Pack (see Sec.~\ref{sec:impl} for particular details) on the prototypical example of Bi$_2$Se$_3$. The $\mathbb{Z}_2$ invariant is calculated for the planes at $k_y = 0$ and $0.5$, with HWCCs calculated along $k_z$, and $k_x$ acting as a pumping parameter. Because Bi$_2$Se$_3$ is symmetric with respect to permutations of the unit cell vectors, this is sufficient to fully determine the topological state~\cite{Soluyanov-PRB11-b}.

\begin{figure}\centering
\includegraphics[width=8.5cm]{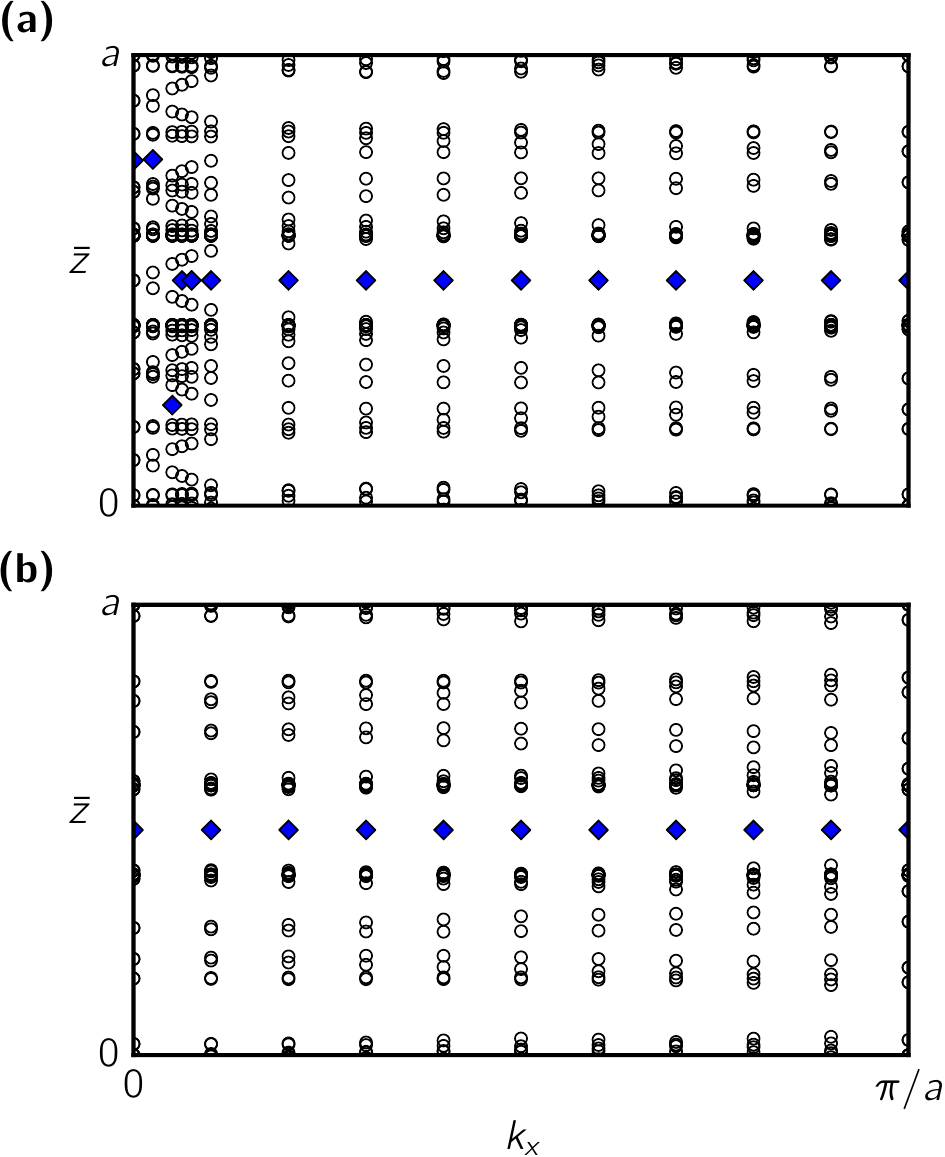}
\caption[]{Evolution of HWCCs (circles) and their largest gap function (blue rhombi) for $k_y = 0$ (a) and $k_y = 0.5$ (b) planes for Bi$_2$Se$_3$. The plane at $k_y=0$ is topologically non-trivial.}
\label{fig:Bi2Se3}
\end{figure}

This calculation was performed with the VASP software package~\cite{VASP}, using the generalized gradient approximation of the PBE~\cite{PBE} type, and the PAW potentials~\cite{PAW1,PAW2} supplied by VASP. The self-consistent calculations were performed with a 12x12x12 k-mesh, an energy cut-off of $300\unit{eV}$ and the experimental lattice parameters~\cite{Perez-ICh99}. The results, shown in Fig.~ \ref{fig:Bi2Se3}, illustrate a non-trivial $\Delta$ for the $k_y=0$ plane and a trivial one for the $k_y=0.5$ plane. Thus, Z2Pack identifies Bi$_2$Se$_3$ as a strong topological insulator, in agreement with previous calculations~\cite{Zhang-NatPhys09} and the parity-eigenvalue argument of Ref.~\cite{Fu-PRB07}. Note that the illustration is provided here for clarity only, and no manual inspection of the plot is needed. The calculation of the $\mathbb{Z}_2$ invariant is fully automated in the code in accord with the method of Ref.~\cite{Soluyanov-PRB11-b}, giving the invariant value as an output. 

\subsection{Crystalline topological materials}
We now discuss the topological phases protected by crystalline symmetries. In principle, any crystalline symmetry can induce a topological classification, however, to date only few such classifications are known~\cite{Fu-PRL11, Parameswaran-NatPhys13, Alexandradinata-PRB16, Alexandradinata-PRL14, Taherinejad-PRB14}. We expect Z2Pack to be most useful for identification of materials with yet unknown topologies, which, in turn, would accelerate the progress towards full classification of possible crystalline topological phases and the corresponding low-energy excitations. The two examples of crystalline topological phases we consider below are those of the mirror-symmetric and four-fold rotational ($C_4$) topological insulators.

\subsubsection{Mirror-symmetric topological phases}
The presence of mirror symmetry in the crystal structure of a material results in the presence of planes in the BZ that are mirror-symmetric. This means that the Bloch states on these planes are eigenstates of a unitary matrix $M$ that describes the action of mirror symmetry. In the presence (absence) of spin-orbit coupling this matrix squares to $-1$ ($1$), due to spin rotation. This means that the eigenvalues of $M$ are $\pm i$ ($\pm 1$) for spinor (scalar) Bloch states on the mirror-symmetric planes.

Thus, one can split the occupied subspace ${\cal H}_\mathrm{set}$ on the mirror-symmetric planes into two subspaces, according to their mirror eigenvalue. For example, with spin-orbit accounted the two projectors $\hat{P}_{\pm i}$ split ${\cal H}_\mathrm{set}$ into ${\cal H}_{\pm i}$ consisting of Bloch states with $M$ eigenvalues $\pm i$ correspondingly. The individual Chern numbers $C_{\pm i}$ are then defined for each of these subspaces. Note that the splitting according to the mirror eigenvalue fixes the individual Chern numbers uniquely, and each of the subspaces has $\mathbb{Z}$ classification. This is different from the case of TR-symmetry considered above, where all even/odd individual Chern numbers were equivalent from the point of view of the $\mathbb{Z}_2$ classification.

The work of Ref.~\cite{Teo-PRB08} introduced the {\it mirror Chern number} defined as $n_{M}= (C_{+i}-C_{-i})/2$. This number can be used as a $\mathbb{Z}$ topological invariant for the systems with TR-symmetry, where $C_i= - C_{-i}$ (assuming the mirror-symmetric plane is also TR-symmetric). In magnetic systems, however, the two individual Chern numbers are not necessarily equal, so that the invariant can be given by two integers $(C_i, C_{-i})$ and the corresponding classification is $\mathbb{Z}\times\mathbb{Z}$.

\begin{figure}\centering
\includegraphics[width=8.5cm]{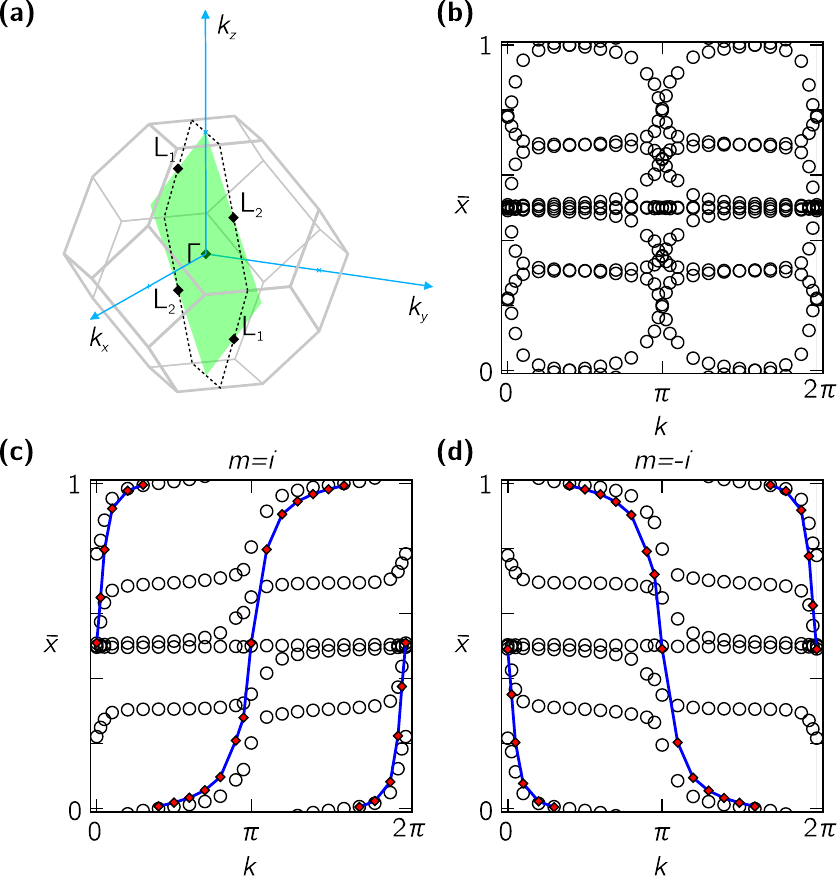}
\caption[]{(a) Brillouin zone of SnTe showing the mirror planes along which the HWCCs are computed. (b) HWCCs in the mirror plane.  (c)-(d) HWCCs (circles) and their sum (rhombi) for the $i$ and $-i$ eigenstates in the mirror plane.}
\label{fig:SnTe}
\end{figure}

A TR-symmetric example of a mirror-symmetric crystalline topological insulator is SnTe, in which the topological phase is protected by the mirror symmetry of its rocksalt structure~\cite{Hsieh-NatComm12,Tanaka-NatPhys12}. The mirror Chern number was predicted to be $n_M=2$ for this material ~\cite{Hsieh-NatComm12}. The individual Chern numbers $C_{+i}$ and $C_{-i}$ are defined on a mirror-invariant plane $(\Gamma L_1 L_2)$ shown in Fig.~\ref{fig:SnTe}(a).

The presence of a topological phase can immediately be inferred by computing HWCCs (running Z2Pack) on the mirror plane. The result of this calculation is shown in Fig.~\ref{fig:SnTe}(b). The absence of a gap in the full HWCCs spectrum, which is a superposition of the HWCCs of both $+i$ and $-i$ mirror eigenstates,  is indeed a strong indicator for the presence of a topological phase.

To compute  the individual Chern numbers $C_{+i}$ and $C_{-i}$ with Z2Pack, it is first necessary to classify each Hamiltonian eigenstate according to the mirror eigenvalues $+i$ or $-i$.
This  is done by computing and diagonalizing at each $k$
the matrix $\bra{\psi_n(\vec{k})} \hat{M} \ket{\psi_m(\vec{k})}$, where $\hat{M}$ is the mirror operator, for all occupied states $\psi_j(\vec{k})$. Using the unitary transformation $U({\vec{k}})$ which diagonalizes this matrix, a set of states with definite mirror eigenvalues is constructed as $|\tilde{\psi}_m(\vec{k})\rangle=\sum_m U_{mn}(\vec{k})|\tilde{\psi}_n(\vec{k})\rangle$. These states are then separated into two groups corresponding to the $\pm i$ eigenvalues, and
Z2Pack is applied to each subspace to compute $C_{+i}=+2$ and $C_{-i}=-2$ as shown in Figs.~\ref{fig:SnTe}~(c) and (d), using the numerical procedure described in Appendix~\ref{app:chern_computation}.

For this illustration, \emph{ab-initio} calculations based on density-functional theory (DFT) \cite{Hohenberg-PR64, Kohn-PR65} were performed, employing the generalized-gradient approximation (GGA) and Perdew-Burke-Ernzerhof exchange-correlation functionals~\cite{PBE} as implemented in the Quantum-ESPRESSO package~\cite{QE-2009}. Spin-orbit effects were accounted for using fully relativistic norm-conserving pseudo-potentials acting on valence electron wave functions, represented in the two-component spinor form~\cite{PhysRevB.71.115106}.  The self-consistent field calculation was performed with a 10$\times$10$\times$10 $k$-mesh, a plane-wave cut-off of $50$~Ry and experimental lattice parameters of Ref.~\cite{SnTe}.

\subsubsection{C$_\text{4}$ topological insulator} \label{sssec:c4}

\begin{figure}\centering
\includegraphics[width=\columnwidth]{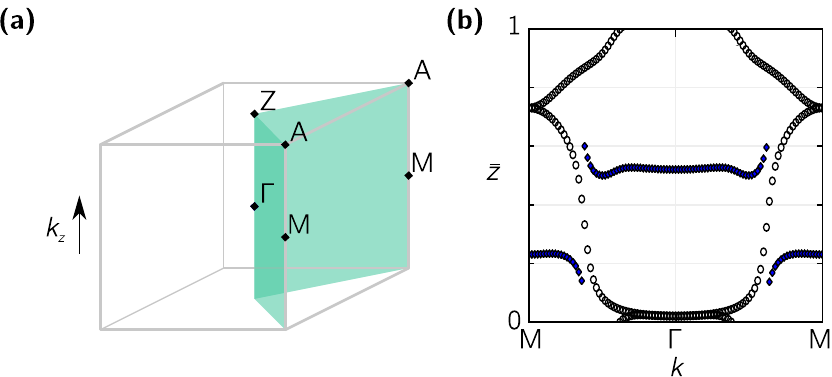}
\caption[]{(a) The Brillouin zone for the model of Ref.~\cite{Fu-PRL11}. The $C_4*\tau$ symmetry maps one leg of the indicated bent plane onto the second. (b) HWCCs (circles) and the largest gap function (rhombi) shown for the $C_4$-symmetric plane of panel~(a).}
\label{fig:c4}
\end{figure}

Certain topological phases are protected by rotation point-group symmetry~\cite{Fu-PRL11, Alexandradinata-PRL14, Alexandradinata-PRB16, Taherinejad-PRB14, Taherinejad-PRL15}. The first model to realize such a phase was proposed  by Fu~\cite{Fu-PRL11}, and it considered spinless fermions with TR-symmetry supplemented with an additional four-fold rotational symmetry $C_4$. The $\mathbb{Z}_2$ classification proposed in Ref.~\cite{Fu-PRL11} arises for bands that belong to two-dimensional representations along the high-symmetry lines $\Gamma$-$Z$ and $A$-$M$ of the BZ shown in Fig.~\ref{fig:c4}~a (the $C_4$-axis is assumed to coincide with the $z$-direction).  For the particular model of Ref.~\cite{Fu-PRL11}, these bands were obtained by considering $p_x$ and $p_y$ orbitals on  a tetragonal crystal lattice with two inequivalent sites, and it was argued that the consideration is also relevant for the bands formed by the $d_{xz}$ and $d_{yz}$ orbitals. 

In real materials such a phase may occur for systems with weak spin-orbit coupling, where the bands near the Fermi level have primarily $p_x$-$p_y$ or $d_{xz}$-$d_{yz}$ character. No real example for this phase was reported to date, and we hope that Z2Pack will encourage people to perform a thorough search of existing materials for the emergence of this novel topology.

The product of a rotation $\hat{C}_4$ and spinless TR-operator $\tau^2=1$ forms a symmetry that is antiunitary and ensures double degeneracy of bands at high-symmetry points due to the constraint $(\hat{C}_4\tau)^2=-1$ for bands of $C_4$ eigenvalue $\pm i$~\cite{Fu-PRL11}. This is analogous to the case of spinful TR-symmetric insulators, where the antiunitary TR-operator $\theta$, subject to the condition $\theta^2=-1$, guaranteed Kramers degeneracy at TR-symmetric momenta in the BZ.
%Analogously to the case of spinful TR-symmetric insulators, where the antiunitary TR-operator $\theta$, subject to the condition $\theta^2=-1$, guaranteed Kramers degeneracy at TR-symmetric momenta in the BZ, here the product of a rotation $\hat{C}_4$ and spinless TR-operator $\tau^2=1$ forms a product symmetry that is antiunitary and ensures double degeneracy of bands at high-symmetry points due to the constraint $(\hat{C}_4\tau)^2=-1$~\cite{Fu-PRL11}. 
To continue this analogy, notice that the $\mathbb{Z}_2$ invariant of the TR-insulators is defined on a plane connecting TR-invariant lines. Thus, the plane for the definition of the invariant in this case should connect lines that are invariant under the product symmetry $C_4*\tau$. %This suggests that this plane consists of two half-planes forming a straight angle.

Such a plane is given by the one shown in Fig.~\ref{fig:c4}, since the double degeneracy of bands is present at the high-symmetry points $\Gamma=(0,0,0)$, $M=(0.5,0.5,0)$, $Z=(0,0,0.5)$ and $A=(0.5,0.5,0.5)$. Due to the $C_4*\tau$ symmetry the HWCCs constructed along lines in $k_z$, corresponding to $\Gamma$-$Z$ and $M$-$A$ directions, are degenerate~\cite{Alexandradinata-PRB16}. Hence, the $\mathbb{Z}_2$ classification on the plane for this model is analogous to that of a single Kramers pair of TR-symmetric insulator on a TR-symmetric plane, as considered in Sec.~\ref{ssec:z2}.

The methods developed above for computing $\mathbb{Z}_2$ topological invariants are applicable in this case. The corresponding HWCCs flow is shown in Fig.~\ref{fig:c4}~b for the topological phase of the model of Ref.~\cite{Fu-PRL11}. From the gapless flow of the HWCCs one can see that the individual Chern numbers $C_{1,2}=\pm 1$ can be assigned to two subspaces mapped onto each other by $\hat{C_4}*\tau$.

The search for a real material candidate for such a phase can proceed as follows. A scalar-relativistic band structure calculation is first performed for compounds of light atoms with small spin-orbit coupling with crystal structures that contain a $C_4$-rotational axis. The spectrum of HWCCs is obtained on a plane shown in Fig.~\ref{fig:c4}~a. Since bands with characters other than $p_x$-$p_z$ and $d_{xz}$-$d_{yz}$ are usually overlapping with these ones, the topological phase can become not immediately visible in the HWCC spectrum.  However, if the HWCCs exhibit strong winding in the spectrum (see Sec.~\ref{ssec:semimetals_intro} for a discussion of this indication of the existence of the topological phase in metals), this might be a hint of the $C_4$-topology buried under HWCCs coming from the non-topological part of the spectrum. In this case, a tight-binding model can be derived, for example using the Wannier90 software package~\cite{wannier90, wannier90_14} or based on symmetry arguments and parameter fitting, that projects the band structure onto the relevant orbitals ($p_x$-$p_y$ or $d_{xz}$-$d_{yz}$), and the WCCs analysis of the tight-binding model will uncover the phase.

Finally, we notice, that the $C_4$-phase illustrated above is only one of the possible phases protected by rotational point group symmetry operations. We refer the reader to the works of Refs.~\cite{Fu-PRL11, Alexandradinata-PRL14, Alexandradinata-PRB16, Taherinejad-PRB14} for a thorough discussion of these phases and the corresponding Wilson loops.

\section{Application to Metals}\label{sec:semimetals}
Here we discuss how the HWCC technique and Z2Pack can be used to identify topological phases in metallic band structures. The discussion is illustrated with the examples of Dirac~\cite{Jeon-NMat14, BiNa3}, type-I and type-II Weyl semimetals~\cite{TaAs, Soluyanov-Nat15, Wang-PRL16}, as well as those of some higher-order topologically protected crossings~\cite{Fang-PRL12, Heikkila-JETP10}.

The Weyl~\cite{Wan-PRB11, Lv-PRX15, Lu-Science15, Xu-Science15, Huang-NatComm15, Huang-Nat15, Weng-PRX15} semimetal phase is characterized by a point-like crossing of bands with a linear spectrum. Topologically, this crossing is characterized by a quantized topological charge, meaning that it is either a source or sink of Berry curvature, depending on its chirality~\cite{Volovik-JETPL87}. As such, Weyl nodes can only form or annihilate in pairs of opposite chirality. The presence of Weyl nodes in the bulk leads to the appearance of Fermi arcs on a surface of the material, connecting the projections of the bulk points of opposite chirality onto the surface. In the presence of a magnetic field, Weyl points exhibit a chiral Landau level~\cite{Adler-PR69, Bell-Jackiw-NC69, Nielsen-PLB83}, which can be a source of the reduced or negative magneto-resistance observed in Weyl semimetals~\cite{Abrikosov-PRB98, Huang-PRX15, Arnold-NatComm16, Yang-ARX15}.

Recently, a new, type-II, kind of Weyl semimetals has been proposed~\cite{Soluyanov-Nat15}. These type-II Weyl nodes appear at the touching points of electron and hole pockets. As a consequence, they are expected to exhibit an anisotropic chiral anomaly~\cite{Soluyanov-Nat15}. That is, the presence or absence of a chiral anomaly depends on the direction of the applied magnetic field, aligned with the electric field. 

In Dirac semimetals~\cite{Wang-PRB12, Liu21022014, BiNa3, Yang-PRB15, Muechler-ARX16}, the nodal points are formed by doubly degenerate bands in systems, symmetric under the product of TR and inversion $P*T$. This symmetry maps a Weyl point onto itself, but with opposite chirality, so that a Dirac point consists of two superimposed Weyl fermions of opposite Chern numbers that are protected from annihilation by lattice symmetries such as $C_n$, where $n>2$. 

Other point-like topological degeneracies can exist in metallic spectra~\cite{Winkler-PRL16, Wieder-PRL16, Bradlyn-Sci16, Zhu-PRX16}. However, all of these phases can be viewed as either a certain superposition of Weyl points, just like insulating topological phases can be viewed as symmetry dictated superpositions of bands with non-zero individual Chern numbers, or as Weyl points, intersected by an additional band. The methodology outlined below is suited to identify such types of topologically protected degeneracies.

\subsection{Generalized topological classification}\label{ssec:semimetals_intro}
In semimetals, the existence of a direct band gap closure means that the manifold formed by the occupied bands is not well-defined. This seemingly contradicts the definition of topologically distinct states because the notion of topologically equivalent states requires them to be adiabatically connectible without a direct band gap closure. This problem is avoided by restricting the consideration to a specific 2D manifold $M$ within the BZ, which does not contain a direct band gap closure. Two states are then considered topologically equivalent on $M$ if they can be adiabatically connected without a direct band gap closure occurring on $M$. Using this generalized notion of topological classification, the methods discussed previously for identifying a general topological phase in insulators can be generalized to the Fermi surfaces of metals.

\begin{figure}\centering
\includegraphics[scale=0.75]{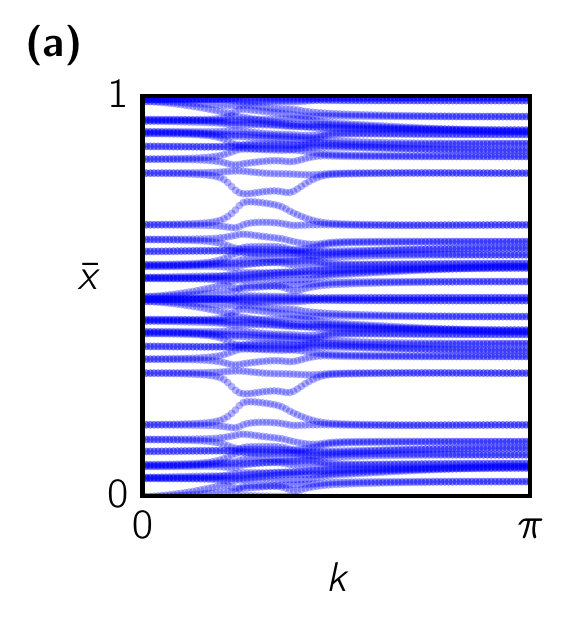}
\includegraphics[scale=0.75]{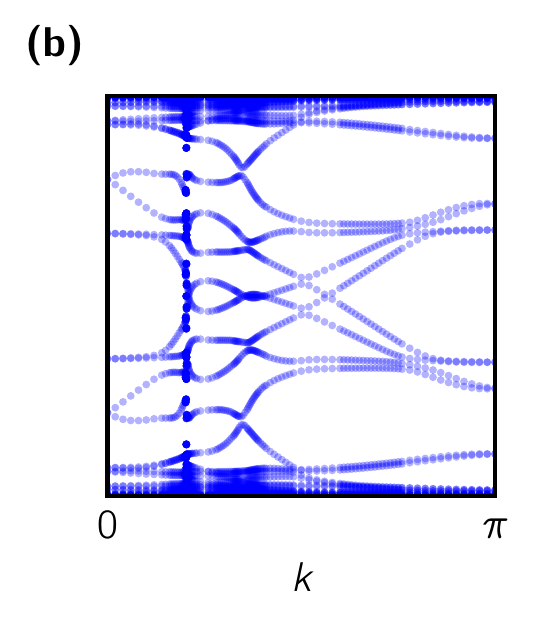}
\caption[]{(a) HWCCs of WTe$_2$ for $k_y=0$ plane~\cite{Soluyanov-Nat15}. A nearly gapless flow of HWCCs indicates the presence of topologically non-trivial features in the band structure close to this plane. (b) HWCCs of MoTe$_2$ for $k_z=0$ plane~\cite{Wang-PRL16}. The presence of a band gap closure is visible as a discontinuity in the HWCCs evolution, and occurs at a momentum, corresponding to the position of a type-II Weyl point in this material.}
\label{fig:semimetal_hints}
\end{figure}

The search for manifolds in metallic BZs, on which the bands have topologically non-trivial features, can be a challenging task. It is aided a lot by the fact that the HWCC technique often reveals hints of non-trivial topology on manifolds, where the band structure is topologically trivial, but close to a phase transition. For example, Fig.~\ref{fig:semimetal_hints}~(a) illustrates HWCCs obtained on the $k_y=0$ plane in WTe$_2$. While the spectrum of HWCCs is gapped, the presence of kinks and the narrow gaps hints at the possible presence of the topologically non-trivial features nearby. Indeed, it was shown in Ref.~\cite{Soluyanov-Nat15} that type-II Weyl points exist in this material in the close vicinity of this plane. 

Furthermore, the presence of a band gap closure within the manifold $M$ is revealed as a divergence in the HWCC calculation, which can be detected by Z2Pack. This is illustrated for another type-II Weyl semimetal, MoTe$_2$~\cite{Wang-PRL16}, in Fig.~\ref{fig:semimetal_hints}, which shows the divergence of the HWCCs lines obtained with Z2Pack on the $k_z=0$ plane in the BZ. This divergence is due to the four type-II Weyl points that appear in this plane owing to the existence of the product symmetry of a $C_2$-rotation around $z$ and TR $\theta$~\cite{route}. The way of finding Weyl points and other degeneracies  in the high symmetry planes by tracking divergencies in the Z2Pack calculation is especially useful for a high-throughput search of topological metals.  

\subsection{Chern and $\mathbb{Z}_2$ invariants in metals}\label{ssec:semimetals_z2}
The first and most straightforward way to examine metallic band structures for the presence of non-trivial topologies is to compute the Chern and $\mathbb{Z}_2$ invariants described above for insulators on planes in the metallic BZ where the bands are gapped. 

For magnetic metals, for example, one should compute Chern numbers on various planes (see Fig.~\ref{fig:cubic_bz}). A change in the value of the Chern invariant when going between the adjacent planes in $k$-space indicates the presence of a topologically protected degeneracy in between the planes, such as a Weyl point. An example of such a material is HgCr\sub{2}Se\sub{4} \cite{Xu-PRL11}.

A similar argument holds for TR-symmetric metals. The $\mathbb{Z}_2$ number can be computed on the standard TR-symmetric planes ($k_i=\{0,0.5\}$), and for inversion-(a)symmetric materials a change in its value suggests the presence of a Dirac point (a pair of Weyl points) in between the planes. We illustrate this point here by showing the use of Z2Pack to identify the Dirac semimetal phase. We use BiNa$_3$~\cite{Wang-PRB12,Liu21022014} as a material example.

\begin{figure}\centering
\includegraphics[width=8.5cm]{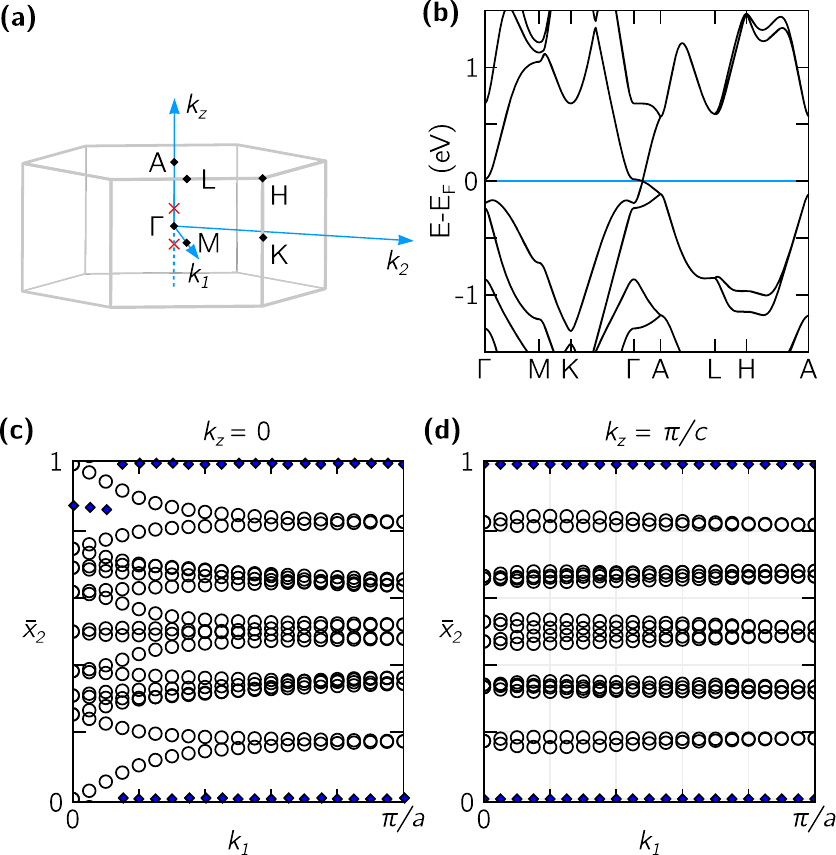}
\caption[]{(a) Brillouin zone of BiNa\sub{3}. The red crosses show the position of the two 3D Dirac points. (b) Band structure of BiNa\sub{3}. (c)-(d) HWCCs (circles) and the largest gap function (rhombi) for the BZ cuts $k_z=0$ and $k_z=\pi/c$.}
\label{fig:BiNa3}
\end{figure}

This material crystallizes in the centrosymmetric hexagonal $P6_3/mmc$ structure and exhibits a band inversion at $\Gamma$, similar to the band inversion observed in the $\mathbb{Z}_2$ topological insulators Bi\sub{2}Se\sub{3} and Bi\sub{2}Te\sub{3}~\cite{Zhang-NatPhys09}. However, this material is not gapped; its Fermi surface consists of two four-fold degenerate Dirac points located at $\vec{k}_d=(0,0,\pm 0.29\frac{\pi}{c})$ (see Fig.~\ref{fig:BiNa3}~(a) and (b)) corresponding to two overlapping Weyl points of opposite chirality.

The Dirac points are formed, since the Weyl points of opposite chirality are protected from gapping each other by a rotational symmetry~\cite{Wang-PRB12, Yang-NatComm14}: the two doublets crossing along $\Gamma$-$A$ belong to different irreducible representations of the little group of the $k$-vectors $\vec{k}=(0,0,u)$ ($C_{6v}$). The topology of such materials can be captured with Z2Pack by computing the $\mathbb{Z}_2$ topological invariants of two TR-invariant planes located above and below the 3D Dirac points as shown in Figs.~\ref{fig:BiNa3}~(c) and (d).

\begin{figure}\centering
\includegraphics[width=8.5cm]{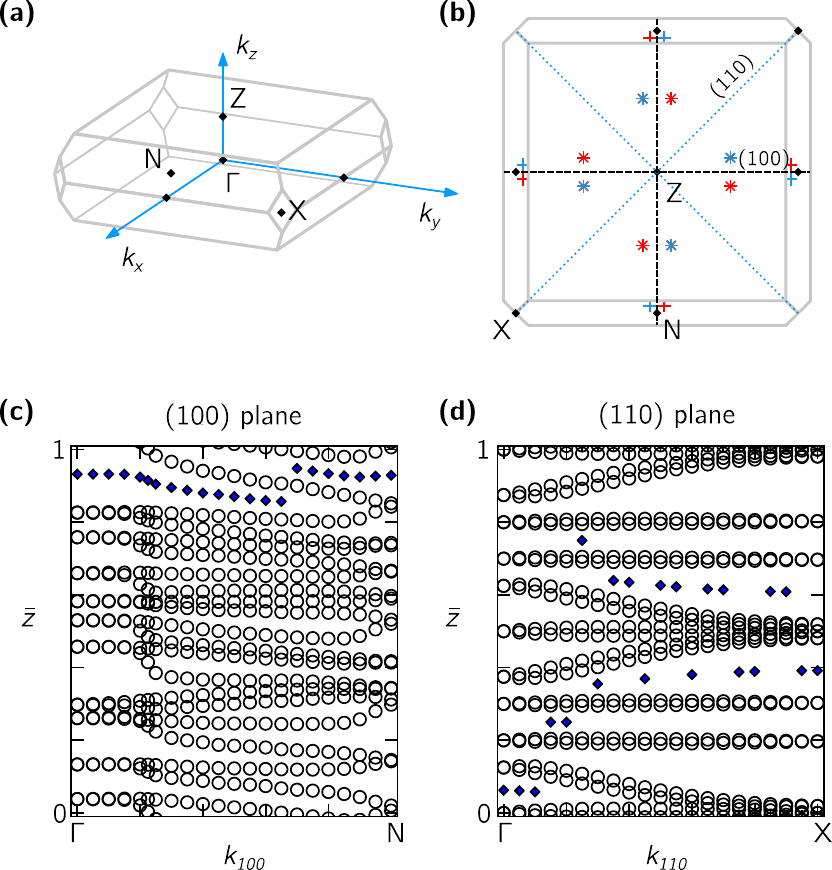}         
\caption[]{(a) Brillouin zone of TaAs. (b) Top view of the Brillouin zone showing the position of the 24 Weyl points with chirality +1 (red) and -1 (blue) (crosses are used for Weyl points in the $k_z=0$ plane, and stars designate Weyl points with $k_z=\pm 0.59k_{\Gamma Z}$).  (c)-(d) HWCCs (circles) and the largest gap function (rhombi) along the [100] mirror plane and [110] glide plane shown in (b).}
\label{fig:TaAs}
\end{figure}

For TR-symmetric materials with no inversion symmetry, the change in the $\mathbb{Z}_2$ value on the TR-symmetric planes in the BZ indicates the presence of Weyl points. An example of such a Weyl semimetal is TaAs~\cite{Weng-PRX15, Lv-PRX15, Huang-NatComm15, Xu-Sci15}. This material crystallizes in the non-centrosymmetric body-centered tetragonal $I4_1 md$ structure. It is a semimetal possessing 24 Weyl points near the Fermi level~\cite{Weng-PRX15, Lv-PRX15, Huang-NatComm15, Xu-Sci15}. Due to TR-symmetry, these 24 points come in 12 pairs. Four symmetry-related pairs reside in the $k_z=0$ plane and the remaining eight are located symmetrically about the [100] and [010] mirror planes at $k_z=\pm 0.59k_{\Gamma Z}$ (marked with crosses and stars in Fig.~\ref{fig:TaAs}~(b)). 

The presence of Weyl points in the TR-symmetric band structure can be identified with Z2Pack. For TaAs, HWCCs were obtained for the $k_y=0$ plane (shown as [100] in Fig.~\ref{fig:TaAs}~(b)) and the result is shown in Fig.~\ref{fig:TaAs}~(c). The corresponding  $\mathbb{Z}_2$ invariant is non-trivial. 
Note, that while in the context of TR-symmetric topological insulators the TR-planes used to compute the topological invariants are those defined by $k_i=\{0, 0.5\}$, the $\mathbb{Z}_2$ invariant is well-defined on any section of the BZ that for any point $\vec{k}$ also contains the point $-\vec{k}$, and that connects lines related by a reciprocal lattice vector.
Thus, one can define a $\mathbb{Z}_2$ invariant on a $k_x=k_y$ plane (shown as [110] in Fig.~\ref{fig:TaAs}~(b)). The corresponding invariant is trivial in TaAs, as illustrated by the evolution of the HWCCs on the half-plane from $\Gamma$ to the TR-invariant point $X$ in Fig.~\ref{fig:TaAs}~(d)~\footnote{The band structure calculations for BiNa$_3$ and TaAs were performed with Quantum-ESPRESSO, following the same methodology as described above for the SnTe calculations. The SCF calculations  were performed with a 10$\times$10$\times$10 $k$-mesh, a plane-wave cut-off of $50$~Ry and experimental lattice parameters taken from Refs.~\cite{BiNa3,TaAs}.}.

Moreover, the cuts of the BZ do not need to be planar to define a $\mathbb{Z}_2$ invariant. As mentioned previously in Sec. \ref{sssec:c4}, the only requirement for a $\mathbb{Z}_2$ invariant is that the surface connects three TR-invariant lines, making the HWCC on these lines doubly degenerate. This allows for a more complete characterization of the $\mathbb{Z}_2$ topology using curved surfaces. 
In fact, identifying topological invariants on such planes allows to guess the possible connectivities of the Fermi arcs on the surfaces of topological semimetals.

\begin{figure}\centering
\includegraphics[height=4.6cm]{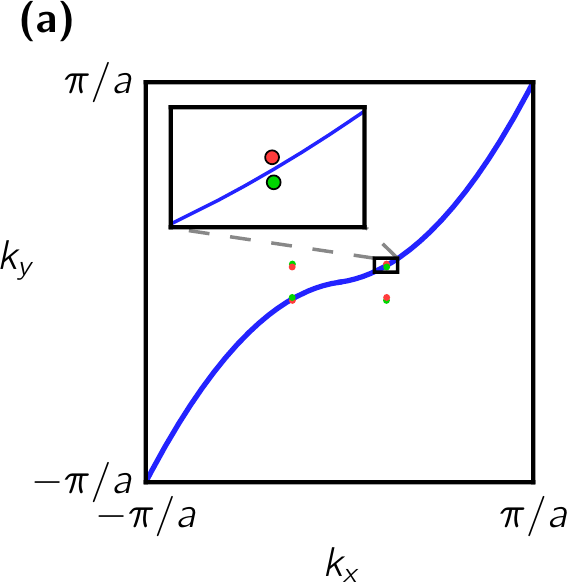}\quad
\includegraphics[height=4.6cm]{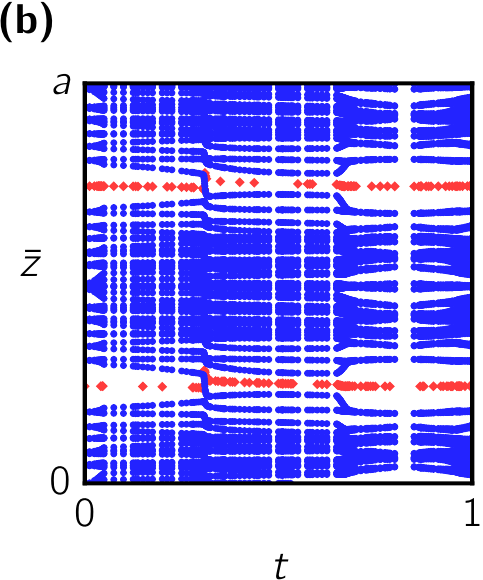} 
\caption[]{ 
(a) Cut through the BZ of WTe\sub{2} at $k_z=0$. The eight WPs are indicated, as well as the curved surface used to calculate the $\mathbb{Z}_2$ invariant (blue line). The surface extends in $k_z$ direction. (b) Evolution of WCC (blue dots) and their largest gaps (red rhombi) along the curved surface indicated in panel (a), exhibiting a non-trivial $\mathbb{Z}_2$ invariant. }
\label{fig:wte2_curved}
\end{figure}

As an example, consider the case of WTe\sub{2}~\cite{Soluyanov-Nat15}. Fig.~\ref{fig:wte2_curved}~(a) illustrates the locations of the type-II Weyl points in the $k_z=0$ plane, of which this material has 8. The $\mathbb{Z}_2$ invariant is trivial on all TR-invariant planes $k_i=0$ and $0.5$ except $k_z=0$, where it is undefined due to the presence of band gap closures. Additionally, the $\mathbb{Z}_2$ invariant was calculated on a TR-invariant curved surface passing in between Weyl points, as illustrated in Fig.~\ref{fig:wte2_curved}~(a), where it was found to be non-trivial. Thus, this topology cannot be characterized from the planes at $k_i=0, 0.5$ alone.

\subsection{Existence and chirality of Weyl Fermions}\label{ssec:semimetals_weyl}
Next we show that the chirality of a Weyl point, as well as other topological point-crossings, can be calculated directly as the Chern number on a closed surface enclosing the node. This provides a simple and reliable way of determining the nature of nodal points.

Let us first calculate the Chern number on a sphere enclosing a Weyl point. The Hamiltonian for an isotropic (upon possible rescaling and rotations) $3$D type-I Weyl point located at the origin is:
\begin{equation}
H(\vec{k}) = \sum_{i=1}^3 k_i \sigma_i 
\end{equation} 
with two energy states $\pm k$. The lower-energy eigenstate ($E=-k$) as a function of the momentum $(k_x, k_y, k_z) = k(\sin\theta \cos\phi, \sin \theta \sin \phi, \cos \theta)$ is:
\begin{equation}
\psi_{\vec{k}}= \frac{1}{\sqrt{2(1- \cos(\theta) )}}
\left(
\begin{array}{cc}
1 - \cos \theta   \\
-\sin \theta e^{i \phi} 
\end{array}
\right)
\end{equation} 
Note that the eigenstate is smooth and well-defined everywhere except for $\theta= 0$. Since 
\begin{equation}
\lim_{\theta\rightarrow 0^+} \frac{\sin\theta}{\sqrt{1 - \cos\theta}}=\sqrt{2}
\end{equation}
the eigenstate takes the following form at the north pole $\vec{k}=(0, 0, 1)$:
\begin{equation}
\psi_{\vec{k} = (0,0, 1) }= 
\left(
\begin{array}{cc}
0   \\
- e^{ i \phi} 
\end{array}
\right)
\end{equation}
which means it is multi-valued. In other words, different values of $\phi$, although describing the same momentum, correspond to unequal values for the wave function. The wave function is thus ill-defined in this gauge at the north pole; this is also the point where the Dirac string between the monopole (Weyl fermion) in the center of the sphere and infinity crosses the Fermi surface. Of course, this is just a gauge choice - by making a gauge transformation we can move the position of the intersection of the Fermi surface with the Dirac string to wherever we want on the sphere. 

The existence of a Weyl fermion can be verified by calculating the flux of Berry curvature through a surface enclosing it. Choosing a sphere of unit radius, the Berry vector potential is given by
\begin{align}
& \vec{\mathcal{A}}(\theta, \phi) = i \bra{\psi_k} \nabla_k \ket{\psi_k} = - \frac{\sin\theta}{2(1 - \cos\theta)} \vec{e}_\phi
\end{align}
The Chern number is thus given by
\begin{align}
C & = \frac{1}{2\pi} \int\limits_{\Omega}\left[\nabla \times \vec{\mathcal{A}}\right] \cdot \text{d}\vec{S} = \\
& = \frac{1}{2\pi} \int\limits_{0}^{2\pi}\text{d}\phi \int\limits_{0}^{\pi}\text{d}\theta\left[ \frac{\partial}{\partial \theta}\left( -\frac{\sin^2 \theta}{2(1-\cos\theta)}\right) \right]=1
\end{align}
Since the Chern number cannot change under smooth deformations of the surface as long as the bands  remain gapped on it, the argument can be generalized to any closed surface. The same is true for adiabatic changes in the Hamiltonian, which cannot change the Chern number on the surface without closing the band gap \textit{on the surface}. Consequently, the Chern number on a sphere can be used to confirm the existence of a Weyl point within the sphere, and determine its chirality. This illustrates that Weyl points can be viewed as a quantized topological charge, which acts as a source or sink of Berry curvature. 

\begin{figure}\centering
\includegraphics[width=0.6\columnwidth]{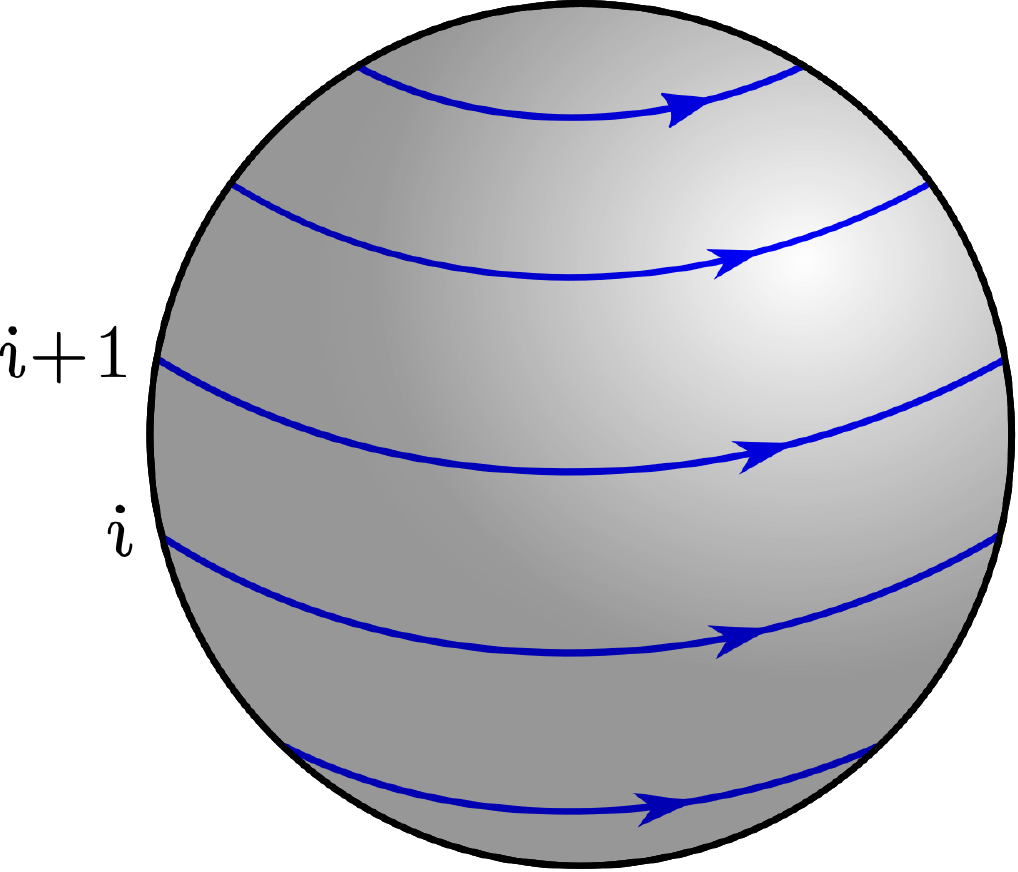}
\caption[]{Loops around a sphere on which WCC are computed. Each loop circles the sphere in mathematically positive direction at a constant azimuthal angle $\theta$. The loops go from the south pole ($\theta=-\pi$) to the north pole ($\theta=0$).}
\label{fig:sphere}
\end{figure}

As described in Sec.~\ref{ssec:chern_hwf} a Chern number on any closed 2D manifold can be calculated by tracking the evolution of the sum of HWCC. In the case of a sphere, the HWCC can be computed on loops around the sphere, as illustrated in Sec. \ref{fig:sphere}. The sum of HWCC is then tracked as a function of the angle $\theta$. For $\theta=-\pi$ and $0$, the loop is just a single point. As a consequence, the bands do not acquire any phase in the parallel transport, and the sum of HWCC must be zero. This ensures the values for $\theta=-\pi$ and $0$ to be the same, even though the two loops are not equivalent lines in the BZ.

\begin{figure}[h]\centering
\includegraphics[width=\columnwidth]{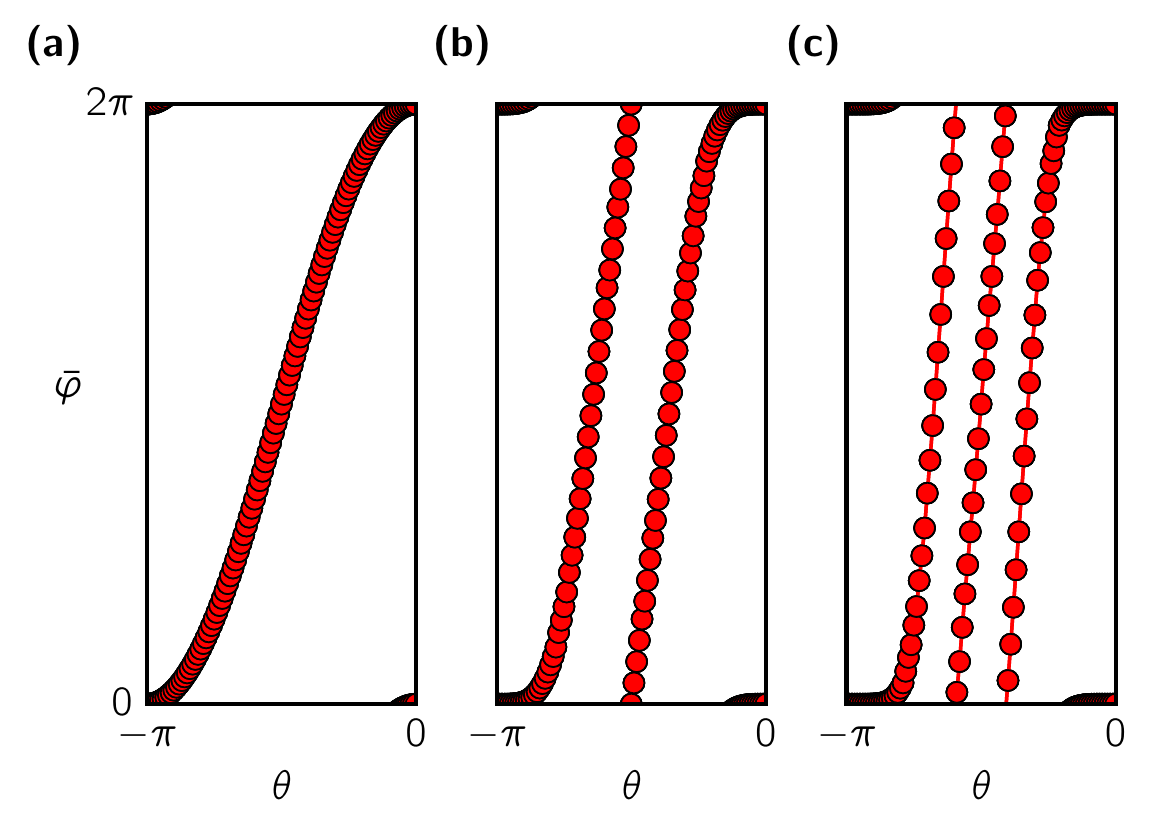}
\caption[]{The evolution of polarization around linear (a, $n$=1), quadratic (b, $n=2$) and cubic (c, $n=3$) touching points in $\mathcal{H}_\text{eff}$.}
\label{fig:weyl_chirality}
\end{figure}

This method also works for topological crossings other than standard Weyl points. For example,
consider an effective Hamiltonian~\cite{Fang-PRL12, Heikkila-JETP10}
\begin{equation}
\mathcal{H}_{\text{eff}}(\vec{k})= \begin{pmatrix}
k_z & (k_-)^n \\
(k_+)^n & -k_z
\end{pmatrix}
\end{equation}
where $n \in \N$ and $k_\pm = k_x \pm i k_y$. The two bands will have an $n$-th order touching point at the origin in the $k_z=0$ plane, while the crossing is linear in the $k_z$ direction. The results for linear, quadratic and cubic touching points are illustrated in Fig.~\ref{fig:weyl_chirality}. Using the method of computing HWCCs on a sphere described above, as implemented in Z2Pack, we found that the Chern number for such an effective Hamiltonian is $C=n$, which agrees with theoretical considerations of Refs.~\cite{Fang-PRL12,Heikkila-JETP10}. A few particular cases are illustrated in Fig.~\ref{fig:weyl_chirality}.

In type-II Weyl points \cite{Soluyanov-Nat15}, the energy spectrum is tilted in such a way that their Fermi surface becomes open. Unlike for type-I Weyl point, where the FS is a sphere around the node, the topological charge of the FS cannot be used to determine its chirality. However, the method described above is not linked to the Fermi surface topology. Indeed, the chirality of a type-II Weyl point can still be determined by considering the lower-lying bands on a surface enclosing the point. In fact, type-II and type-I Weyl points of the same chirality can be adiabatically connected, which means the associated Chern number of the surface must be the same. On the other hand, this means that the type of a Weyl point cannot be determined by means of calculating topological invariants.

\begin{figure}[h]\centering
\includegraphics[width=\columnwidth]{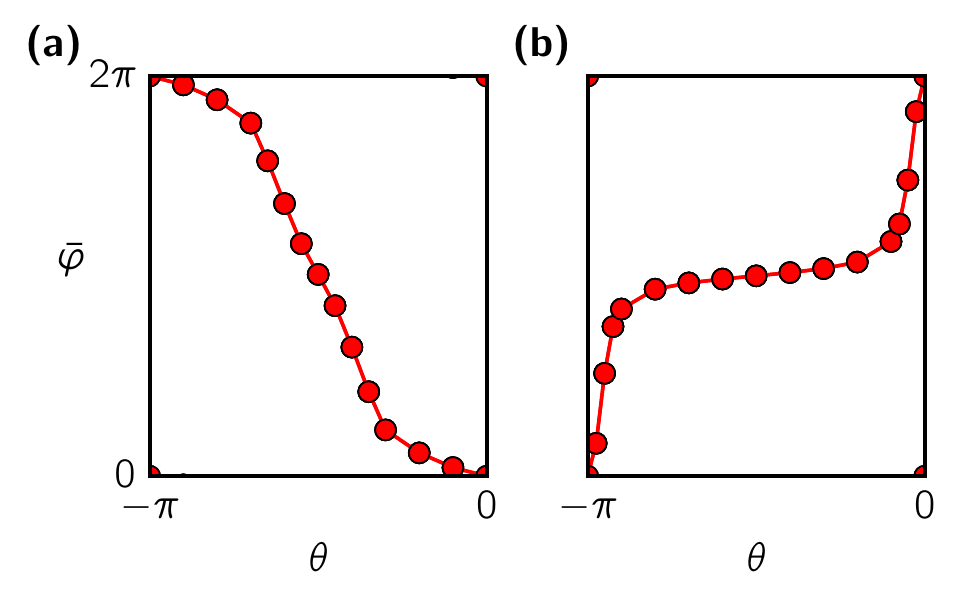}
\caption[]{Change in charge polarization on a sphere surrounding Weyl points in WTe\sub{2}, indicating their chirality. (a) Sphere of radius $r=0.005$ around a Weyl point of negative chirality at $\vec{k}=(0.1203, 0.05232, 0.0)$. (b) Sphere of radius $r=0.005$ around a Weyl point of positive chirality at $\vec{k}=(0.1211, 0.02887, 0.0)$.}
\label{fig:wte_chirality}
\end{figure}
Fig.~\ref{fig:wte_chirality} shows the evolution of polarization on a sphere around two of the type-II Weyl points in WTe\sub{2}. The HWCCs were calculated from a tight-binding model derived from first-principles, with full spin-orbit coupling.

\subsection{Dirac Semimetals}

In Dirac semimetals, the nodal point consists of two degenerate Weyl nodes of opposite chirality, mapped into each other by the product of time-reversal and parity. An additional symmetry is required to keep the two Weyl nodes from gapping. Since each of the two Weyl nodes contributes an individual Chern number $\pm 1$, we expect to see a gapless flow in the HWCC evolution on a sphere enclosing a Dirac point. 

We exemplify this by studying Cd\sub{3}As\sub{2}, which was recently shown to be a Dirac semimetal \cite{Wang-PRB13, Jeon-NMat14}. The modified four-band $\vec{k}\cdot \vec{p}$ Hamiltonian used to study this material is given by \cite{Wang-PRB13, Soluyanov-Bernevig-unpublished}\footnote{The Hamiltonian used here differs from that of Ref.\cite{Wang-PRB13} in that terms up to third order are explicitly taken into account.}
\begin{align}
\mathcal{H}(\vec{k}) = & \varepsilon_0 (\vec{k}) \mathbb{I} \otimes \mathbb{I} + M(\vec{k}) \tau_z \otimes \mathbb{I} + \\\nonumber & +\left[ A k_x + B k_x^3 + F k_x k_y^3 + G k_z^2 k_x \right] \tau_x \otimes \sigma_z - \\\nonumber & - \left[ A k_y + B k_y^3 + F k_y k_x^2 + G k_z^2 k_y \right] \tau_y \otimes \mathbb{I} + \\\nonumber & + n_1 k_z \left( k_x^2 - k_y^2 \right) \tau_x \otimes \sigma_x + n_2 k_x k_y k_z \tau_x \otimes \sigma_y.
\end{align}
where 
\begin{equation}
M(\vec{k}) = m_0 + \sqrt{m_3^2 + m_1 k_z^2} + m_2 (k_x^2 + k_y^2),
\end{equation}
and
\begin{equation}
\varepsilon_0 (\vec{k}) = c_0 + c_1 k_z^2 + c_2 (k_x^2 + k_y^2).
\end{equation}

The parameters from Ref. \cite{Jeon-NMat14}
\begin{align}
& m_0 = -0.06 ~\unit{eV} & m_1 = 96 ~\unit{eV^2 \AA^2} \\\nonumber
& m_2 = 18 ~\unit{eV \AA^2} & m_3 = 0.05 ~\unit{eV} \\\nonumber
& c_0 = -0.219 ~\unit{eV} & c_1 = -30 ~\unit{eV\AA^2}\\\nonumber
& c_2 = -16 ~\unit{eV \AA^2} & A = 2.75 ~\unit{eV \AA},
\end{align}
were used, and different values for $B, F, G, n_1, n_2$ were studied to investigate the properties of the Dirac point when higher-order corrections are included.

\begin{figure}\centering
\vspace{1em}
\includegraphics[height=3.8cm]{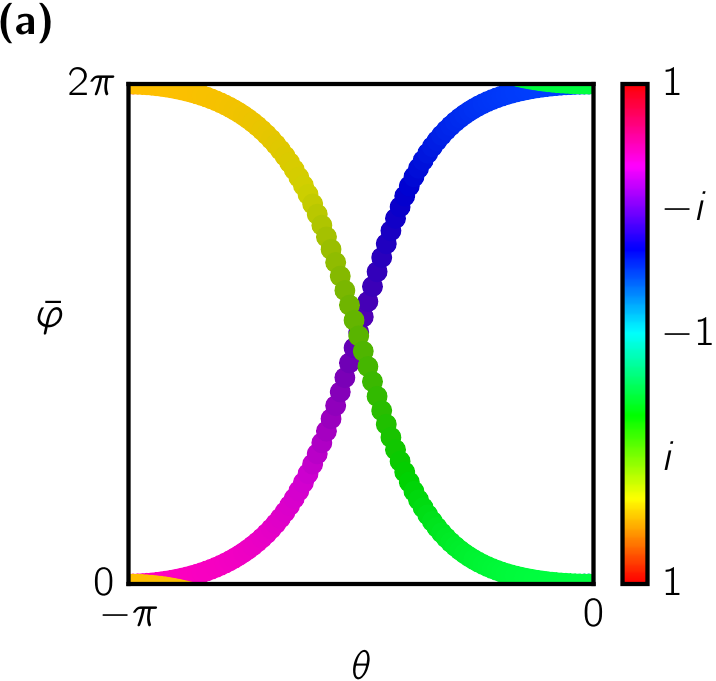}\quad
\includegraphics[height=3.8cm]{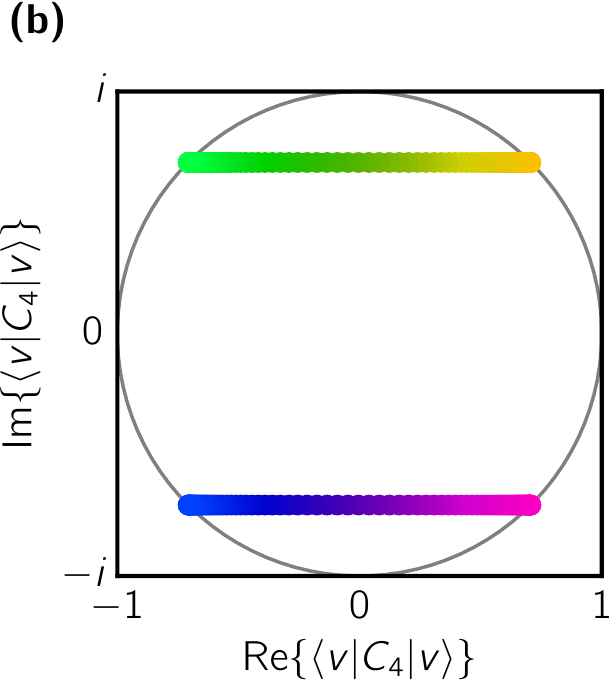}\\\vspace{1em}
\includegraphics[height=3.8cm]{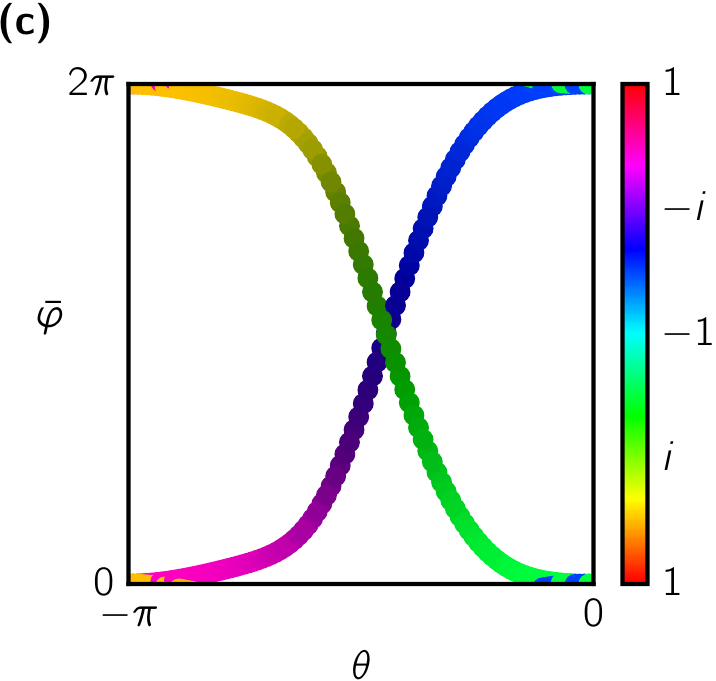}\quad
\includegraphics[height=3.8cm]{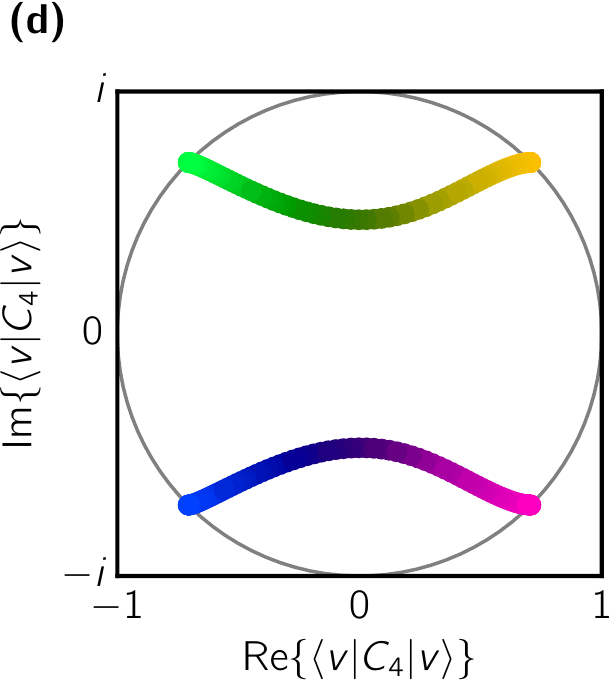}
\caption[]{(a) WCC evolution on a sphere of radius $r=0.001 \unit{\AA^{-1}}$ enclosing one of the two Dirac points in Cd\sub{3}As\sub{2} with $n_1=n_2=0$. The WCC are colored according to the $C_4$ expectation values of the corresponding Wilson loop eigenstates. The expectation values are mapped on the complex plane in panel (b). (c), (d)  WCC evolution and $C_4$ expectation values on the same sphere around one of the Dirac points, for $n_1=n_2=10^6 \unit{eV \AA^3}$}
\label{fig:Cd3As2_C4}
\end{figure}

As expected, the HWCC evolution on a sphere surrounding one of the two Dirac points appears gapless (see Fig.~\ref{fig:Cd3As2_C4}~(a), (c)). However, from this consideration alone it is not clear whether these HWCC indeed form a crossing, or whether there may be some small gap. Unlike the case of the $\mathbb{Z}_2$ classification described above, there is {\it a priori} no symmetry which enforces the HWCC  to be degenerate at some $\theta$. To prove the existence of a Dirac node, then, it is necessary to consider the effect of the symmetry stabilizing the Dirac fermion on the HWCC. This can be done by calculating the symmetry expectation value of the Wilson loop eigenstates $v$ which correspond to the given HWCC (see Appendix \ref{app:symmetry_exp_value} for details). As can be seen in Fig.~\ref{fig:Cd3As2_C4}~(b) and (d), the eigenstates have different $C_4$ expectation values at the point where the HWCC cross. This means small perturbations cannot gap the HWCC flow. 

Furthermore, in the case of $n_1=n_2=0$, the two HWCC belong to two different subspaces $V_{\pm}$ spanned by the eigenstates with eigenvalues $\{ e^{+ i \pi / 4}$, $e^{+ i 3 \pi / 4}\}$ and $\{ e^{+ i \pi / 4}$, $e^{+ i 3 \pi / 4}\}$, respectively. The individual Chern numbers $c_\pm = \mp 1$ corresponding to each of these subspaces reveal the presence of Weyl points of opposite chirality. When higher order terms are included in the Hamiltonian (see Fig.~\ref{fig:Cd3As2_C4}~(c)-(d)), the Wilson loop eigenstates no longer belong to one of the two subspaces $V_\pm$, since they are mixed by the $\tau_x \otimes \sigma_x$ and $\tau_x \otimes \sigma_y$ terms. However, the mixing term, being quadratic in $k_x, k_y$, becomes vanishingly small, when the radius of the sphere surrounding the Dirac point is taken to be small. 

We thus conclude that the presence of topological nodal points, comprised of several overlapping Weyl points, like Dirac point, can be revealed by the flow of Wannier charge centers, provided one tracks the expectation values of the symmetry that protects the Weyls from annihilating for the corresponding eigenstates of the non-Abelian Berry connection (Wilson loop) to make sure that they are distinct at the crossing point in the WCC spectrum. In this case the WCC spectrum is gapless and the topological phase is proven.

\section{Numerical Implementation} \label{sec:impl}

Here we outline the numerical implementation of the methodology described in the previous sections. 
The method of calculating (individual) Chern numbers and $\mathbb{Z}_2$ invariants on different manifolds in the BZ is implemented in the {\tt Z2Pack} code package, which is an open-source Python~\cite{Python} module. The code and documentation, including tutorials and examples, are available in the supplementary material to this paper. Updated versions of the code and documentation will be made available online (code: \url{https://pypi.python.org/pypi/z2pack/}, documentation: \url{http://z2pack.ethz.ch}).

One-dimensional maximally localized hybrid Wannier charge centers are computed directly from the overlap matrices as defined in Appendix~\ref{ssec:multiband_wcc}. The Chern and $\mathbb{Z}_2$ invariants can be automatically extracted from the WCC, by using the methods described in Appendix~\ref{app:chern_computation} and~\ref{app:z2_computation}. The numerical calculations are performed with help of the \texttt{numpy}~\cite{numpy} and \texttt{scipy}~\cite{scipy} packages.

Z2Pack is compatible with any method or software, which can provide the overlap matrices or eigenstates for a given path of k-points. Tools for computing the overlap matrices for tight-binding and $\vec{k} \cdot \vec{p}$ models are included in the module. For first-principles computations, an interface to the Wannier90~\cite{wannier90, wannier90_14} code is provided, and the overlap matrices are computed by the first-principles codes that support Wannier90, making Z2Pack compatible with any such code. For example, widely used VASP~\cite{VASP}, Quantum Espresso~\cite{QE-2009} and ABINIT~\cite{Gonze-CPC09, Gonze-ZK05} codes can be straightforwardly used with Z2Pack.

Furthermore, Z2Pack features a rich set of convergence criteria to ensure the correct evaluation of the topological indices. This is especially important because of the quantized nature of the topological invariants, making it impossible to approximate their value iteratively. In all but the most delicate cases \footnote{If the direct band gap becomes very small, the WCC tend to move very quickly. It is then necessary to use a more stringent convergence criteria. However, it is still possible to achieve automatic convergence with Z2Pack.}, Z2Pack will converge automatically using only the provided default parameters. This makes the code ideally suited for high-throughput applications by minimizing the need for manual intervention.

Finally, Z2Pack provides methods for plotting the results. Figures showing the WCC and their largest gap (such as in Fig.~\ref{fig:Bi2Se3}), the sum of WCC (see Fig.~\ref{fig:weyl_chirality}), and the WCC colored according to a symmetry expectation value (see Fig.~\ref{fig:Cd3As2_C4}) can be produced. The plotting functions are based on the \texttt{matplotlib}~\cite{matplotlib} package, and their appearance can be fully customized.

\section{Conclusions}\label{sec:conclusion}

We introduced and enumerated the known approaches to identifying topological states in both insulators and semimetals and provided an easy-to use package for the evaluation of topological invariants such as Chern numbers, $Z_2$ invariants, mirror Chern numbers, and semimetal monopole charges. The approach is based on the calculation of hybrid WCCs from the overlap matrices constructed with Bloch states. We showed how the method can be used to classify part of the knowns symmetry-protected topological states of non-interacting systems, based on the notion of individual Chern numbers. The proposed scheme is suited for high-throughput search for materials with non-trivial topology and can point to materials that have yet undiscovered non-trivial topologies.

We also presented a numerical implementation of the method in the Z2Pack software package.Examples were provided for materials with various topologies. For insulators, the Chern and $\mathbb{Z}_2$ topological phases were illustrated, as well as some crystalline TIs. For semimetals, we illustrated approaches for classifying and identifying topological nodal points. Generalizations to nodal lines are straightforward. 

\section*{Acknowledgements}
We would like to thank G. Winkler, M. K\"onz and D. Rodic for useful discussions. D.G., A.A.S. and M.T. were supported by the Swiss National Science Foundation through  the  National  Competence  Centers  in  Research MARVEL and QSIT. G.A. and O.V.Y. acknowledge support by the NCCR Marvel and the ERC Starting grant ``TopoMat'' (Grant No. 306504). D.V. was supported by NSF Grant DMR-1408838. B.A.B. was supported by Department of Energy DE-SC0016239,NSF EAGER Award NOA - AWD1004957, Simons Investigator Award, ONR - N00014-14-1-0330, ARO MURI W911NF-12-1-0461, NSF-MRSEC DMR-1420541, Packard Foundation and Schmidt Fund for Innovative Research

\bibliography{paper}

\appendix
\section{Numerical computation of the Chern number}\label{app:chern_computation}

The straightforward way to compute the Chern number is to integrate the gauge-invariant Berry curvature over the (section of) BZ. A numerical calculation using the hybrid WCCs is also possible. In this case the Chern number is computed by evaluating electronic polarization at discrete points $k_y=k_i \in [0, \frac{2 \pi}{a}]$.
\begin{equation}
P_e(k_i) = e \left(\sum_n\bar{x}_n (k_i) \mod a_x\right)
\end{equation}
Because the polarization is defined only modulo $e\cdot a$, the same is true for the change in polarization, whose possible values are 
\begin{equation}
\Delta P_{e, i} = P_e(k_{i+1}) - P_e(k_i) + l (e\cdot a)
\end{equation}
for $l \in \mathbb{Z}$. Assuming that the $k$-points $k_i$ are dense enough such that the true change in polarization is less than $\frac{e \cdot a}{2}$ between any two steps, the correct choice of $l$ is the one that minimizes the absolute value of $\Delta P_{e, i}$. The Chern number is then given by
\begin{align}
C &= \frac{1}{e a} \sum_i \Delta P_{e, i} \\&=  \frac{1}{e a} \sum_i \min_{k \in \Z} \left[ P_{e, i + 1} - P_{e, i} + k(e\cdot a)\right]_\text{abs},
\end{align}
where $\min\left[ \cdot \right]_\text{abs}$ denotes the minimum with respect to the absolute value.

\section{Numerical computation of the $\mathbb{Z}_2$ invariant}\label{app:z2_computation}

Here we describe how the $\mathbb{Z}_2$ invariant is calculated for a given set of WCC 
\begin{equation}
\{\bar{x}^i_n:=\bar{x}_n(k_i), n \in \{1, ..., N\}, i \in {1, ..., M}\},
\end{equation}
where we assume the WCC to be normalized to $[0, 1)$.
We define $g^i := g(k_i)$ as the largest gap between any two WCC $x_n^i$. That is, $g^i$ is such that the distance to the nearest WCC 
\begin{equation}
\min_{n} d(g^i, x_n^i)
\end{equation}
is maximized, where $d$ is the periodic distance. This distance can be numerically evaluated as
\begin{equation}
d(x, y) = \min\left(|1 + x - y| \mod 1, |1 - x + y| \mod 1\right).
\end{equation}

For each step $i \rightarrow i + 1$, the number $n_i$ of WCC $\bar{x}^{i+1}_n$ for which
\begin{equation}
\min(g^i, g^{i+1}) \leq \bar{x}_n^{i+1} < \max(g^i, g^{i+1})
\end{equation}
is counted. This is equivalent to the number of crossings between the largest gap and the WCC between $k_i$ and $k_{i+1}$. Thus, the $\mathbb{Z}_2$ invariant is given by 
\begin{equation}
\Delta = \left(\sum_{i=1}^{M - 1} n_i\right) \mod 2.
\end{equation}

\section{Computation of 1D maximally localized hybrid Wannier charge centers}\label{app:hwcc_computation}
\subsection{Single-band systems}
Numerically, the Berry phase of a single-band system can be computed as a product of overlaps $\langle u_{k_i} | u_{k_{i+1}}\rangle$ along a string of $k$-points $k_i$ going across the BZ.
\begin{gather}
\prod\limits_{i=0}^{N-1} \langle u_{k_i}|u_{k_{i+1}}\rangle = c_N \cdot e^{-i \tilde{\varphi}_B (N)} \xrightarrow[N\rightarrow \infty]{} e^{-i \varphi_B}; ~~~~ c_N \in \R\\
\Rightarrow \tilde{\varphi}_B(N) = -\arg \left[  \prod\limits_{i=0}^{N-1} \langle u_{k_i}|u_{k_{i+1}}\rangle  \right ]
\end{gather}
This can be cast in terms of the parallel transport of the Bloch state across the BZ. For this, in going from $k_i$ to $k_{i + 1}$, the state $|u_{k_{i + 1}}\rangle$ is rotated such that it is parallel to $|u_{k_i}\rangle$, so that their overlap is real and positive: 
\begin{gather}
|\tilde{u}_{k_{i + 1}}\rangle = e^{-i \arg[\langle u_{k_i}|u_{k_{i+1}}\rangle]} |u_{k_{i + 1}}\rangle \\ \label{eqn:condition_parallel}
\Rightarrow \langle u_{k_i}|\tilde{u}_{k_{i + 1}}\rangle \in \R^+
\end{gather}
Doing this procedure along the closed loop from $k_0$ to $k_N$ (see Appendix~\ref{App-C} for the explicit expressions for overlaps), a total phase $\tilde{\varphi}_B(N)$ is picked up, which converges to the exact Berry phase for large $N$.
\subsection{Multi-band systems}
\label{ssec:multiband_wcc}
The same principle of rotating the states along a closed path keeping them parallel to each other in consecutive steps is applied when more than one band is present. However, the overlap is now defined as a matrix
\begin{equation} 
M_{m,n}^{(k_i, k_{i+1})} = \langle u_{m, k_i}| u_{n,k_{i+1}}\rangle.
\label{eqn:overlap_matrix}
\end{equation}
The states at $k_{i+1}$ must be rotated in such a way that the resulting overlap matrix is hermitian. From a singular value decomposition $M = V\Sigma W^\dagger$, this rotation can be obtained as $WV^\dagger$~\cite{Marzari-PRB97}. Along a closed path, the states pick up a non-Abelian phase~\cite{Wilczek-PRL84}
\begin{equation}
\Lambda = W_{n-1}V_{n-1}^\dagger  \dots  W_0V_0^\dagger
\label{eqn:lambda}
\end{equation}
whose eigenvalues $\lambda_i$ are connected to WCCs by
\begin{equation}
\bar{x}_i = - \frac{\arg(\lambda_i)}{2\pi}.
\end{equation}
Note that this construction gives the WCC normalized to $[0, 1)$. 
\section{Phase shift originating from atomic positions in tight-binding models}\label{App-C}
Tight-binding models are defined as a system of orbitals $\ket{\phi_\alpha},~\alpha \in \{1, \dots, N\}$, localized at positions $\vec{t}_\alpha$ within the unit cell, and a set of on-site energies $E_\alpha$, as well as hoppings between the orbitals. A hopping between orbitals $\ket{\phi_\alpha}$ and $\ket{\phi_\beta}$, located in unit cells specified by lattice vectors $\vec{R}_\alpha$ and $\vec{R}_\beta$ correspondingly, is in general given by a complex number $s \in \mathbb{C}$. A hopping matrix can be introduced with entries at $(\alpha, \beta)$ and $(\beta, \alpha)$
\begin{equation}
A(\alpha, \beta, s) =  (s e^{i \vec{k}\cdot \vec{T_{\alpha,\beta}}} \delta_{\alpha, i}\delta_{\beta, j})_{i, j} + h.c.
\end{equation}
where $\vec{T}_{\alpha, \beta} = \vec{R}_\alpha - \vec{R}_\beta$ is the vector connecting the positions of the two orbitals.

We make a gauge convention such that the total Hamiltonian matrix is given by
\begin{equation}
H(\vec{k})= \text{diag}(E_1,\dots,E_N) + \sum\limits_{i} A(\alpha_i, \beta_i, s_i)
\end{equation}
This guarantees that $H({\bf k}+{\bf G})=H({\bf k})$, where ${\bf G}$ is a reciprocal lattice vector. 

Given the Hamiltonian, its eigenvectors $\ket{\psi_n(\vec{k})} = \sum\limits_\alpha c^n_\alpha(\vec{k}) \ket{\phi_\alpha}$  can be computed. An overlap matrix element in the adopted convention is given by
\begin{equation}
M_{m,n}^{(\vec{k}, \vec{k}+\vec{b})} = \sum\limits_\alpha^\text{occ.}c_\alpha^{m}(\vec{k})^* c_\alpha^n(\vec{k + b})e^{-i \vec{b}\cdot \vec{t_\alpha}},
\end{equation}
assuming the orbitals $\ket{\phi_\alpha}$ are perfectly localized at $\vec{t}_\alpha$.

Unlike the Hamiltonian itself, the overlap matrices depend on the orbital positions $\vec{t}_\alpha$, which act as a phase factor. However, both the symmetry of the system and its spectrum are determined by the Hamiltonian alone. It is thus possible to adiabatically move the orbital positions to the origin without changing the topology of the system, provided the hoppings are kept unchanged and the space group of the system is symmorphic. In this case (but not for non-symmorphic systems) the phase factor originating from the orbital positions in the unit cell, can be ignored when computing topological invariants (but not electronic polarization, which is not quantized in general).

\section{Projector expression for the individual Chern numbers}
\label{app:individual_chern}
Here we show that the total Chern number associated with an isolated set of bands is decomposed into the sum of individual Chern numbers as defined in Sec.~\ref{ssec:individual_chern}.

Let $H$ be a Hilbert space spanned by the bands $\left\{\ket{i}:= \ket{u_{\vec{k}, i}}, i \in \left\{1, ..., N\right\} \right\}$, which are defined on a smooth and closed 2D manifold $M$.
The family of projectors 
\begin{equation}
P_\vec{k} = \sum\limits_{i=1}^N \ketbra{i}{i}
\end{equation}
onto $H$, as well as the families of projectors 
\begin{equation}
P_\vec{k}^{(i)} = \ketbra{i}{i}
\end{equation} 
onto the individual bands are all assumed to be smooth on the manifold.

The Chern number associated with these bands is then given by (Eq.~\ref{eqn:chern_projector})
\begin{equation}
C = \frac{i}{2\pi} \int\limits_M \tr \left\{P_\vec{k} \left[\partial_{k_1} P_\vec{k}, \partial_{k_2} P_\vec{k}\right] \right\} \mathrm{d}k_1 \wedge \mathrm{d}k_2,
\end{equation}
 
By using $\tr\left\{A\right\} = \sum\limits_n \bra{n} A \ket{n}$ and $\braket{i}{n} = \delta_{i, n}$ we find
\begin{equation}
C = \frac{i}{2\pi} \int\limits_M \sum\limits_{i=1}^N \bra{i} \left[\partial_{k_1} P_\vec{k}, \partial_{k_2} P_\vec{k}\right] \ket{i} \mathrm{d}k_1 \wedge \mathrm{d}k_2
\end{equation}

The summand can be simplified as follows:
\begin{align}
&\bra{i} \left[ \partial_{k_1} P_\vec{k}, \partial_{k_2} P_\vec{k} \right] \ket{i} \\\nonumber
=& \sum\limits_{m, n=1}^N\bra{i} \Big[ \partial_{k_1} \ketbra{n}{n}, \partial_{k_2} \ketbra{m}{m} \Big] \ket{i} \\\nonumber 
=& \sum\limits_{m, n=1}^N \bra{i} \Big( \partial_{k_1} \ketbra{n}{n} \partial_{k_2} \ketbra{m}{m} - \partial_{k_2} \ketbra{m}{m} \partial_{k_1} \ketbra{n}{n} \Big) \ket{i}\\\nonumber
=& \sum\limits_{n=1}^N \Big( \bra{i} \partial_{k_1}\ketbra{n}{n} \partial_{k_2} \ket{i} - \bra{i} \partial_{k_2}\ketbra{n}{n} \partial_{k_1} \ket{i} \Big)\label{eqn:sum_n}\\\nonumber
=& ~i \sum\limits_{n=1}^N 2 \Im \Big[ \bra{i} \partial_{k_1} \ketbra{n}{n} \partial_{k_2} \ket{i} \Big]
\end{align}
For cases where $n\neq i$
\begin{align}
&\Im \Big[ \bra{i} \partial_{k_1} \ketbra{n}{n} \partial_{k_2} \ket{i} \Big] \\\nonumber
&= \Im \Big[ \underbrace{\braket{i}{\partial_{k_1} n}}_{=0} \braket{n}{\partial_{k_2} i} + \underbrace{\braket{i}{n}}_{=0} \partial_{k_1} \bra{n}\partial_{k_2} \ket{i} \Big] = 0,
\end{align}
where we used the fact that 
\begin{equation}
\braket{i}{n} = \delta_{in},
\end{equation}
and thus
\begin{equation}
0 = \partial_{k_1} \braket{i}{n} = \braket{\partial_{k_1} i}{n} + \braket{i}{\partial_{k_1} n} = 2 \braket{i}{\partial_{k_1}n}.
\end{equation}
From this, it follows that 
\begin{equation}
\bra{i} \left[ \partial_{k_1} P_\vec{k}, \partial_{k_2} P_\vec{k} \right] \ket{i} = \bra{i} \left[ \partial_{k_1} P_\vec{k}^{(i)}, \partial_{k_2} P_\vec{k}^{(i)} \right] \ket{i}
\end{equation}
and hence 
\begin{align}
C &= \frac{i}{2\pi} \int\limits_M \sum\limits_{i=1}^N \bra{i} \left[\partial_{k_1} P_\vec{k}, \partial_{k_2} P_\vec{k}\right] \ket{i} \mathrm{d}k_1 \wedge \mathrm{d}k_2 \label{eqn:ctot_single_band}\\\nonumber
&= \sum\limits_{i=1}^N \frac{i}{2\pi} \int\limits_M  \bra{i} \left[\partial_{k_1} P_\vec{k}^{(i)}, \partial_{k_2} P_\vec{k}^{(i)}\right] \ket{i} \mathrm{d}k_1 \wedge \mathrm{d}k_2\\\nonumber
&= \sum\limits_{i=1}^N \frac{i}{2\pi} \int\limits_M \tr \left\{P_\vec{k}^{(i)} \left[\partial_{k_1} P_\vec{k}^{(i)} , \partial_{k_2} P_\vec{k}^{(i)} \right] \right\} \mathrm{d}k_1 \wedge \mathrm{d}k_2 \\\nonumber
&= \sum\limits_{i=1}^N c_i
\end{align}
This proves Eq. \ref{ctot} for the special case where the $H_i$ each consist of a single band. Using this special case, the result can be generalized to any splitting of the Hilbert space
\begin{equation}
H = \bigoplus_{i=1}^N H_i
\end{equation}
Let $\left\{ \ket{i_j}, i \in \left\{1, ..., N\right\}, j \in \left\{1, ..., N_i\right\} \right\}$ be a basis of $H$ such that $\left\{ \ket{i_j}, \left\{j \in 1, ..., N_i\right\}  \right\}$ is a basis of $H_i$ for all $i$. It is well-known that such a basis always exists. Applying Eq. \ref{eqn:ctot_single_band} first on $H$ and then on each of the $H_i$, we get
\begin{equation}
C = \sum\limits_{i=1}^N\sum\limits_{j=1}^{N_j} c_{i_j} = \sum\limits_{i=1}^N c_i,
\end{equation}
thus proving Eq. \ref{ctot} for a general splitting of the Hilbert space.

\section{Splitting of the Hilbert space into subspaces according to their symmetry behaviour} \label{app:symmetry_op}
Here we discuss how the Hilbert space can be split into subspaces according to their symmetry, for the cases of unitary and antiunitary symmetry operations.

In the case of a unitary symmetry operation, the Hilbert space can uniquely be split into subspaces which correspond to the eigenspaces of the symmetry operator. For the case of an antiunitary symmetry operation $A$, the same is true for the eigenspaces of the squared symmetry operator $\Gamma = A^2$, which is again unitary. For any $\omega \neq 1$, the eigenstates of $\Gamma$ come in pairs, with eigenvalues $\omega$ and $\omega^*$ \cite{Wigner-JMP60-a}. This creates a special case for $\omega = -1$, where $\omega = \omega^*$ is true. It is then possible to split the eigenspace in two in such a way that for each such pair, only one state is contained in each subspace. However, this splitting is not unique because the two states in a pair may be switched. Consequently, the individual Chern number of the two subspaces is meaningful only if the symmetry relates the individual Chern numbers of the two states in a pair. An important example of such a symmetry is time-reversal, where the two states in a pair must have opposite individual Chern numbers, and thus a switching of states can change the individual Chern number of the two subspaces only by an even number.

\section{Calculation of symmetry expectation values of the Wilson loop eigenstates}\label{app:symmetry_exp_value}
Here we discuss how the symmetry expectation values are calculated for the Wilson loop eigenstates.

The Wilson loop (Eq.~\ref{eqn:proj_occ}, \ref{eqn:wilson}) can be written as a product of overlap matrices (Eq.~\ref{eqn:overlap_matrix})
\begin{gather}
W(\mathcal{C}) = \prod\limits_{i=0}^{L - 1} \sum\limits_{j=1}^{N_\text{occ.}} \ket{u_{j, \vec{k}_i}}\bra{u_{j, \vec{k}_i}} \\\nonumber
= \sum_{j_1, j_2} \ket{u_{j_1, \vec{k}_0}} \left( \prod\limits_{i=0}^{L-2} M^{\vec{k_i}, \vec{k_{i + 1}}} \right)_{j_1, j_2} \bra{u_{j_2, \vec{k}_{L - 1}}}.
\end{gather}
Since the loop $\mathcal{C}$ is closed and thus $\ket{u_{j, \vec{k}_0}} = \ket{u_{j, \vec{k}_{L-1}}}$, the Wilson loop in the basis $\{u_{j, \vec{k_0}}\}_j$ is simply given by the product of overlap matrices
\begin{equation}
W = \prod\limits_{i=0}^{L-2} M^{\vec{k}_i, \vec{k}_{i+1}}\vspace{0.5em}
\end{equation}
and its eigenstates $\ket{v_i}$, fulfilling $W \ket{v_i} = \lambda_n \ket{v_i}$ can be calculated. 

Knowing the symmetry representation $C$ in the basis of the Hamiltonian (that is, the basis in which the $\ket{u_{j, \vec{k}_0}}$ are written), the symmetry expectation values of $\ket{v_i}$ can be calculated by
\begin{equation}
\bra{v_i} \hat{C} \ket{v_i} = v_i^T S^T C S v_i
\end{equation}
where $S$ is the basis-transformation matrix which contains $\ket{u_{j, \vec{k}_0}}$ as its columns.

In the limit of large $L$, $W = \Lambda^\dagger$, where $\Lambda$ is defined as in Eq.~\ref{eqn:lambda}. This means the eigenvalues $\lambda_n$ are related to the (normalized) WCC by 
\begin{equation}
\bar{x}_i = \frac{\arg(\lambda_i)}{2 \pi},
\end{equation}
and the symmetry expectation values can thus be assigned to  corresponding WCCs.

\end{document}